\documentclass[twocolumn]{aastex631}
\usepackage{graphicx}
\usepackage{amsmath}

\begin{document}

\title{Secondary standards in the UKIRT faint standard fields}

\correspondingauthor{Marek G\'{o}rski}
\email{mgorski@camk.edu.pl}

\author[0000-0002-3125-9088]{Marek G\'{o}rski}
\affiliation{Nicolaus Copernicus Astronomical Center, Polish Academy of Sciences, Bartycka 18, 00-716, Warsaw, Poland}

\author[0000-0002-9443-4138]{Grzegorz Pietrzy\'{n}ski}
\affiliation{Nicolaus Copernicus Astronomical Center, Polish Academy of Sciences, Bartycka 18, 00-716, Warsaw, Poland}

\author[0000-0002-0136-0046]{Paulina Karczmarek}
\affiliation{Nicolaus Copernicus Astronomical Center, Polish Academy of Sciences, Bartycka 18, 00-716, Warsaw, Poland}


\author[0000-0003-0594-9138]{Gergely Hajdu}
\affiliation{Nicolaus Copernicus Astronomical Center, Polish Academy of Sciences, Bartycka 18, 00-716, Warsaw, Poland}

\author[0009-0001-6501-1080]{Miros{\l}aw Kicia}
\affiliation{Nicolaus Copernicus Astronomical Center, Polish Academy of Sciences, Bartycka 18, 00-716, Warsaw, Poland}

\author[0000-0003-0109-5833]{Miko{\l}aj Ka{\l}uszy\'{n}ski}
\affiliation{Nicolaus Copernicus Astronomical Center, Polish Academy of Sciences, Bartycka 18, 00-716, Warsaw, Poland}

\author[0000-0001-6976-180X]{Joseph R. Eimer}
\affiliation{Johns Hopkins University, Department of Physics and Astronomy, 3701 San Martin Drive, Baltimore, MD 21218, USA}

\author{Stephen A. Smee}
\affiliation{Johns Hopkins University, Department of Physics and Astronomy, 3701 San Martin Drive, Baltimore, MD 21218, USA}

\author[0000-0003-1515-6107]{Bart{\l}omiej Zgirski}
\affiliation{Universidad de Concepci\'{o}n, Departamento de Astronom\'{i}a, Casilla 160-C, Concepci\'{o}n, Chile}

\author[0000-0002-1662-5756]{Piotr Wielg\'{o}rski}
\affiliation{Nicolaus Copernicus Astronomical Center, Polish Academy of Sciences, Bartycka 18, 00-716, Warsaw, Poland}

\author[0000-0003-2335-2060]{Weronika Narloch}
\affiliation{Nicolaus Copernicus Astronomical Center, Polish Academy of Sciences, Bartycka 18, 00-716, Warsaw, Poland}


\begin{abstract}
We present precise J- and K-band photometric measurements for 128 near-infrared secondary standard stars, located in the 19 UKIRT/MKO primary faint standard fields. The data were collected over more than 50 nights, covering a decade of observations between 2008 and 2018 at the ESO La Silla Observatory, using the New Technology Telescope (NTT) equipped with the SOFI NIR camera. Presented magnitudes are calibrated onto the MKO photometric system. The J- and K-band magnitudes range from 10 to 15.8 mag, with median values of $\tilde{J}$ = 13.5 and $\tilde{K}$ = 13 mag. The selection process ensured high photometric quality, with a precision better than 0.01 mag for all stars. The catalog excludes stars with close neighbors, high proper motion, or variable stars. Using these fields for standardization can improve the precision and accuracy of photometric calibrations without increasing the observational time cost.
\end{abstract}



\section{Introduction} \label{sec:intro}
The era of modern near-infrared (NIR) astronomical observations began in the 1960s with the development of highly sensitive PbS photometers. Unlike the previously used InSb detectors, the new cell could be cooled with liquid nitrogen to 77 K. This improvement reduced and stabilized the thermal radiation of the instrument, enabling brightness measurements up to 5~\textmu m. At the same time, the photometric system was expanded with J-, K-, L- and M-bands, centered at approximately 1.3, 2.2, 3.6 and 5.0~\textmu m, respectively \citep{Johnson1962}. 

Soon after, \citet{Johnson1966} presented a list of J- and K-band measurements of 653 bright stars, setting up the first list of NIR standards. It is worth noting that NIR photometry was obtained in two different observatories -- Catalina Station of the Lunar and Planetary Laboratory of the University of Arizona and in the Tonantzintla Observatory in Mexico -- using two different photometers. The absolute calibration of this system is anchored to Vega, with a V-band magnitude of 0.03 and $V – J$ and $V – K$ colors of 0.01, which consequently yields a $J – K$ color of 0.

In the next decade, selected observatories began near-infrared observations with photometers based on PbS and InSb detectors. Often, each observatory had its own in-house set of standard stars, composed of a subsample of objects from the \citet{Johnson1966} list and extended with additional bright stars. This list of observatory-specific standards was anchored using different approaches. The first list used in the South African Astronomical Observatory (SAAO) was published by \citet{Glass1974} and was standardized each night with a subsample of roughly 20 stars from the \citet{Johnson1966} list. This list was later improved and expanded by \citet{Carter1990}, who adjusted the zero points of the J- and K-band magnitudes so that the locus of the $V-K$ and $V-J$ relations against $B-V$ passed through the origin. 
In 1978, \citeauthor{Frogel1978} published a list of 22 standard stars used by the Caltech/Tololo (CIT) observatories, which was complemented with fainter stars by \citet{Elias1982}. The zero points of the CIT system were established by adopting 0.00 magnitudes and colors for Vega. A different approach was used at the ESO La Silla observatory, where a set of 87 stars was calibrated to match the Vega 0.00 magnitude in the V-band, but the NIR zero points were shifted to match the solar energy distribution \citep{Engels1981,Wamsteker1981}. The Mount Stromlo Observatory (MSO) system \citep{Jones1982} was tied with the fundamental standard (HR3314) to the \citet{Glass1974} measurements for this star. The MSO system provided the basis for the development of the Anglo-Australian Observatory (AAO) standard list \citep{Allen1983}, which was additionally composed of stars from the \citet{Glass1974} and \citet{Frogel1978} lists. The AAO system was later refined as the Mount Stromlo and Siding Springs Observatory (MSSSO) system \citep{McGregor1994}.

Although the different approaches used for the zero-point calibration of the described systems introduced only systematic shifts, it was already clear that comparing the brightness of stars between these systems is more complex. Despite the fact that all systems were based on the \citet{Johnson1966} list, it could not be used as a common reference due to its insufficient accuracy.
Furthermore, the lists of standards for the northern and southern hemispheres remained separate, and only a limited number of comparison stars were available. Moreover, those early lists contained systematic errors, and variable stars were present.
At different observatories, the filters used had varying characteristics; they differed in effective wavelength, half-power width, and, in principle, tended to be too broad, often including atmospheric lines. This effect was further amplified by the unique characteristics of atmospheric transparency at different observatories. Finally, the detectors exhibited different spectral responses and deviations from linearity. Even if specific color-based transformations between systems were established, they would fail for particular stars with strong absorption lines or those that were heavily reddened. Despite the problems described above, it was possible to establish color transformations between the CIT and AAO systems \citep{Elias1983}, as well as between the ESO and SAAO/AAO systems \citep{Bouchet1991}.

In the 1990s, the introduction of NIR CCD arrays and the increasing size of telescope mirrors enabled measurements of fainter objects. However, it also revealed the need for fainter standards, as the existing list contained objects that were too bright for modern detectors, which became saturated under normal observing conditions. The SAAO list was extended with standards of brightness up to 10 magnitudes in the K-band by \citet{Carter1995}, and \citet{Bouchet1991} introduced fainter stars to the ESO list. The CIT \citep{Elias1982} system was the basis for a fainter standard list maintained at the 3.8 m UK Infrared Telescope (UKIRT) \citep{Casali1992}, which was later adopted for calibration of 86 stars in the northern hemisphere of the ARNICA system \citep{Hunt1998}. \citet{Elias1982} list was also used as the basis for a new faint NIR standard system of the Las Campanas Observatory (LCO, or NICMOS) \citep{Persson1998}. The UKIRT fundamental list was refined by \citet{Hawarden2001} and in its final version it consists of 83 standard stars with K-band magnitudes ranging from 9.5 to 15 mag. Although the list was based on the early-type stars of the CIT list, the magnitudes and colors of these stars were corrected, which established the UKIRT ''natural'' system. In the following years those standards were used extensively at different observatories, including the ESO La Silla and Paranal observatories. 

Soon after, significant progress was achieved in terms of standardization and homogenization of photometric systems. \citet{Tokunaga2002} specified a new set of NIR filters designed to maximize throughput while simultaneously minimizing sensitivity to atmospheric water vapor, reducing background noise, and improving photometric transformations and color dependence in the extinction coefficient. All NIR telescopes at the Mauna Kea Observatory and many other around the world were equipped with these new filter system (including UKIRT, NASA Infrared Telescope Facility, Canada-France-Hawaii Telescope, Keck, Gemini, Subaru, Anglo-Australian Observatory, Nordic Optical Telescope, Osservatorio Astrofisico di Arcetri, Telescopio Nazionale Galileo, and ESO), and it was recommended as the preferred NIR photometric system by the IAU Working Group on Infrared Photometry.
The compilation of standard stars, calibrated in the new MKO system was prepared by \citet{Leggett2006}, and is composed of 79 standards from the UKIRT list of \citet{Hawarden2001} and 42 stars from the LCO/NICMOS list \citep{Persson1998}. 

At the same time, large NIR surveys began operating, covering large parts of the sky, including DENIS \citep{Fouque2000A&AS}, UKIDSS \citep{Lawrence2007} and 2MASS \citep{Cohen2003,Skrutskie2006}. This development opened the possibility of measuring the brightness of the program stars relative to the catalog magnitude of a given survey, provided that the catalog stars were present in the same field and the photometric systems were sufficiently similar. Based on the UKIDSS and VISTA surveys, \citet{Leggett2020} presented a list of 81 standard stars with a median K-band brightness of 17.5 mag, dedicated to  8-m class telescopes, and future extremely large 30- to 40-m class telescopes.

Notably, the existing lists of standards mainly consist of stars that are too bright and tend to saturate detectors. Additionally, standard stars with precise measurements are sparsely distributed (typically one per field), which either requires significant overhead to achieve the desired standardization precision or reduces precision to maintain low observational overheads.

These limitations were acknowledged and addressed by the authors during research conducted as part of the Araucaria Project, which crucially depends on the precision and accuracy of NIR photometry. The Araucaria Project \citep{ara2023} is an international collaboration dedicated to improving the cosmic distance scale using primary distance indicators, including Cepheids \citep{pietrzyn2002a, Gieren2005, Zgirski2017}, the tip of the red giant branch \citep{Gorski2018}, carbon stars \citep{Zgirski2021}, RR Lyrae stars \citep{Karczmarek2017}, and late-type eclipsing binaries \citep{Pietrzyn2019}.

As part of this project, we have collected a substantial volume of high-quality data, which we have recently decided to publish and make available to the scientific community \citep{Karczmarek2021}. In this paper, we present a list of secondary standard stars, calibrated and selected based on 10 years of NIR observations, located in 19 UKIRT faint standards fields. 

The paper is organized as follows. In Section \ref{sec:obs} we describe the NIR observations and instrumental calibrations. Photometry and standarization are detailed in Sections \ref{sec:phot} and \ref{sec:standarization}, respectively. In Section \ref{sec:selection}, we outline the selection criteria. The results are discussed in Section \ref{sec:results}. Appendix \ref{app:obs_log} presents the observing log and detailed data for all standard fields analyzed.

\section{Observations and instrumental calibrations} \label{sec:obs}
 Data were collected over more than 50 nights, covering a decade of observations between 2008 and 2018 at the ESO La Silla Observatory, using the New Technology Telescope (NTT) equipped with the SOFI NIR camera \citep{Moorwood1998}. Using the Large Field (LF) mode of the instrument, its field of view was $4.9' \times 4.9'$ with a pixel scale of $0.288''\,\mathrm{pix}^{-1}$. The observations were obtained during 18 observing runs carried out under multiple ESO proposals dedicated to the study of Cepheid's and eclipsing binaries in the Magellanic Clouds

In addition to the program stars, each night a set of 5 to 14 standard stars from the list of \citet{Hawarden2001} was observed to secure the calibration of the measurements into the standard system. In this paper only observations of the specific fields containing standard stars are analyzed. Table \ref{tab:obs_log} in Appendix \ref{app:obs_log} reports on which standards were observed each night. 

Observations were performed using the dithering technique, where five consecutive exposures of a given field were shifted in both axes by $20''$, relative to the previous position (\texttt{SEQ.OFFSETX.LIST: "0 20 0 -40 0"}, \texttt{SEQ.OFFSETY.LIST: "0 20 -40 0 40"}). The subintegration times (DIT) ranged from 1.2 to 10 seconds, depending on the brightness of the standard star and seeing conditions, with 2, 3, 4 or 6 NDITs per one dither position.

Instrumental calibrations were typically performed shortly after the observations were made; however, over the course of the decade, the calibrations adhered to the procedures outlined in \citet{Pietrzyn2002b}. Basic routines included bad pixel correction, cosmic rays removal, dark correction and flat fielding, incorporating the \texttt{special\_flat.cl} \textsc{IRAF} procedure provided by ESO on the SOFI website. In denser fields, sky subtraction was performed with a two-step process
using the \texttt{XDIMSUM} \textsc{IRAF} package. In the first step, the sky map was obtained by taking the median of all dithered positions. The preliminary map was then subtracted from each individual image, detected stars were masked, and a second background map was calculated. Finally, all images were corrected for the sky background and stacked into the final image. For sparse fields, only one-step sky subtraction was used.

\section{Photometry} \label{sec:phot}


Photometry was performed individually for all FITS files in the J- and K- bands, separately for each field, using a dedicated pipeline based on the \textsc{DAOPHOT II} software package \citep{Stetson1987}. Measurements were obtained for a set of six apertures, ranging in diameter from $1''$ to $6''$, with the sky background estimated within a concentric annulus of $7''$ inner and $10''$ outer diameter. Although the results presented in this paper are based entirely on aperture photometry, PSF photometry was also performed for denser fields (FS014, FS017, FS035, FS121) to subtract neighboring stars and assess the accuracy of aperture photometry. In all cases, the corrections derived from PSF photometry remained below the reported photometric errors.

As a result, a set of photometric files corresponding to each FITS file was obtained, effectively creating a list of magnitudes for a given field at a specific observation date (epoch). The instrumental coordinates were transformed into the ICRS WCS coordinates by cross-referencing with the Gaia DR3 catalog \citep{Gaia2023}. We note that coordinates presented in this paper were finally transformed to epoch 2000 using the \textsc{AstroPy} package \citep{astropy2022}, including proper motions if available from the Gaia query.

The resulting lists of stars for individual epochs (observing dates), along with their coordinates, instrumental magnitudes, and corresponding errors, were cross-matched, creating a time series of instrumental magnitudes for all stars in the field.

In order to bring instrumental measurements from different epochs to the same reference level and to identify stars with excess variability, we performed differential photometry. This procedure does not serve to obtain the final calibrated magnitudes, but rather to detect stars with additional noise that may remain hidden in the standardized photometry. Differential photometry avoids the extra uncertainty of the absolute calibration step and thus provides a more sensitive diagnostic of the photometric stability. A key aspect of this procedure is to correctly select comparison stars in the same field and remove objects that show excess noise. For this purpose, we developed an iterative method comprising three main steps.

\medskip
\noindent Step 1: Initial Estimation of RMS Using a Single Comparison Star.
For each target star, we selected a single comparison star - typically the primary standard in the field. The differential magnitude was calculated for each epoch, and the root mean square (RMS) of these differences was calculated. This RMS was then compared with the formal photometric error reported by the DAOPHOT for the target star.

DAOPHOT computes the formal error by accounting for the photon noise of the star, the noise from the sky background, and the detector's readout noise. However, in our case, the contribution of the readout noise is not accurately included because we did not provided a map of the number of dither positions stacked within a single pixel. As a result, this leads to an underestimation of the error, particularly for fainter objects.

In contrast, the calculated RMS includes a broader set of noise sources: contributions from photon noise, sky background noise, and the detector readout noise of both the target star and the comparison stars. Additionally, it incorporates other noise sources, such as residual errors from flat-fielding and instrumental calibration, detector edge effects, and, if present, intrinsic stellar variability.

\medskip
\noindent Step 2: Fitting the RMS – Error Relation.
To approximate the relationship between the calculated RMS and the formal DAOPHOT error, we fit a function of the following form:

\begin{equation}
\label{eq:err}
f(x) = \log_{10} \left( 10^{2x} + a \right) + b,
\end{equation}
where $x$ is the DAOPHOT error and $a$ and $b$ are free parameters of the function, corresponding to the additional noise. The function \eqref{eq:err} is fitted with a custom procedure. From the entire sample of stars, five stars were randomly selected and using a \texttt{curve\_fit} procedure from the \textsc{SciPy} package \citep{scipy2020}, the parameters
$a$ and $b$ were determined. This process is repeated multiple times (typically 10 times the number of stars, but no more than 1000 repetitions), and the final parameter values are taken as the median of $a$ and $b$ \footnote{The described procedure is similar to RANSAC; however, in RANSAC, the optimal model is selected based on the maximum number of data points that fit the model (e.g., by excluding outliers using 3-sigma clipping). In our approach, evaluating the optimal parameters as the median is sufficient for the intended purpose, and applying the full RANSAC procedure would require modeling the residuals. }.

\medskip
\noindent Step 3: Selection of Comparison Stars.
Using the fitted function \ref{eq:err} we selected new comparison stars for each target star. A star is qualified as a valid comparison star if it does not exceed the corresponding value of the fitted function by more than 0.01 mag, and if its formal DAOPHOT error is below 0.04 mag. The differential magnitude correction is calculated separately for each comparison star, and the final magnitude is obtained as the weighted average, with weights based on the RMS from Step 1. 

\medskip
Steps 2 and 3 were repeated (II iteration), using the newly calculated RMS for both comparison star selection and weighting. In practice, this final iteration had a marginal effect on the corrected magnitudes, but was retained for consistency.

\begin{figure}[ht!]
\plotone{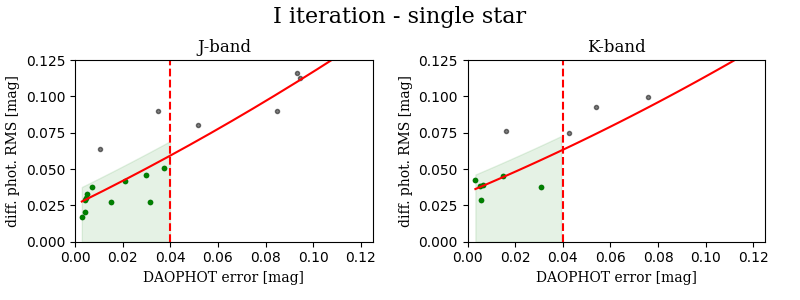}
\plotone{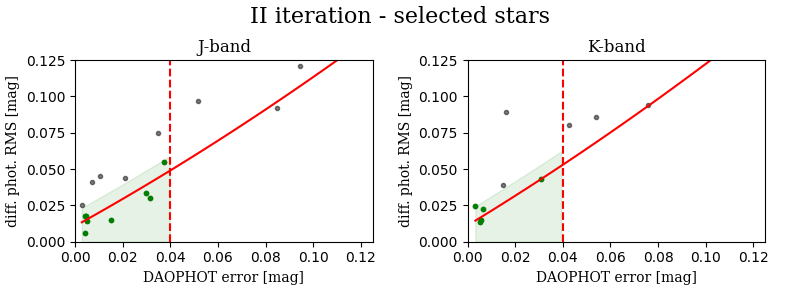}
\caption{RMS versus the average DAOPHOT error for all stars in the exemplary field FS001. The upper panels display the results of the first iteration of the differential correction, while the lower panels present the results of the second iteration. The red solid lines represent the fitted models (eq. \ref{eq:err}). Comparison stars (green points) were selected if their RMS value did not exceed the corresponding value of the fitted function by more than 0.01 mag, and if its formal DAOPHOT error was below 0.04 mag (dashed red vertical line). The green shaded area indicates the region where both criteria are satisfied.
\label{fig:comparison_stars}}
\end{figure}

Figure \ref{fig:comparison_stars} shows the calculated RMS versus the average DAOPHOT error for all stars in the exemplary field FS001.

The calculated RMS will be used to select secondary standards in Section \ref{sec:selection}, and differentially corrected instrumental magnitude time series are saved for further examination and analysis.

\section{standardization} \label{sec:standarization}
\begin{figure}[ht!]
\plotone{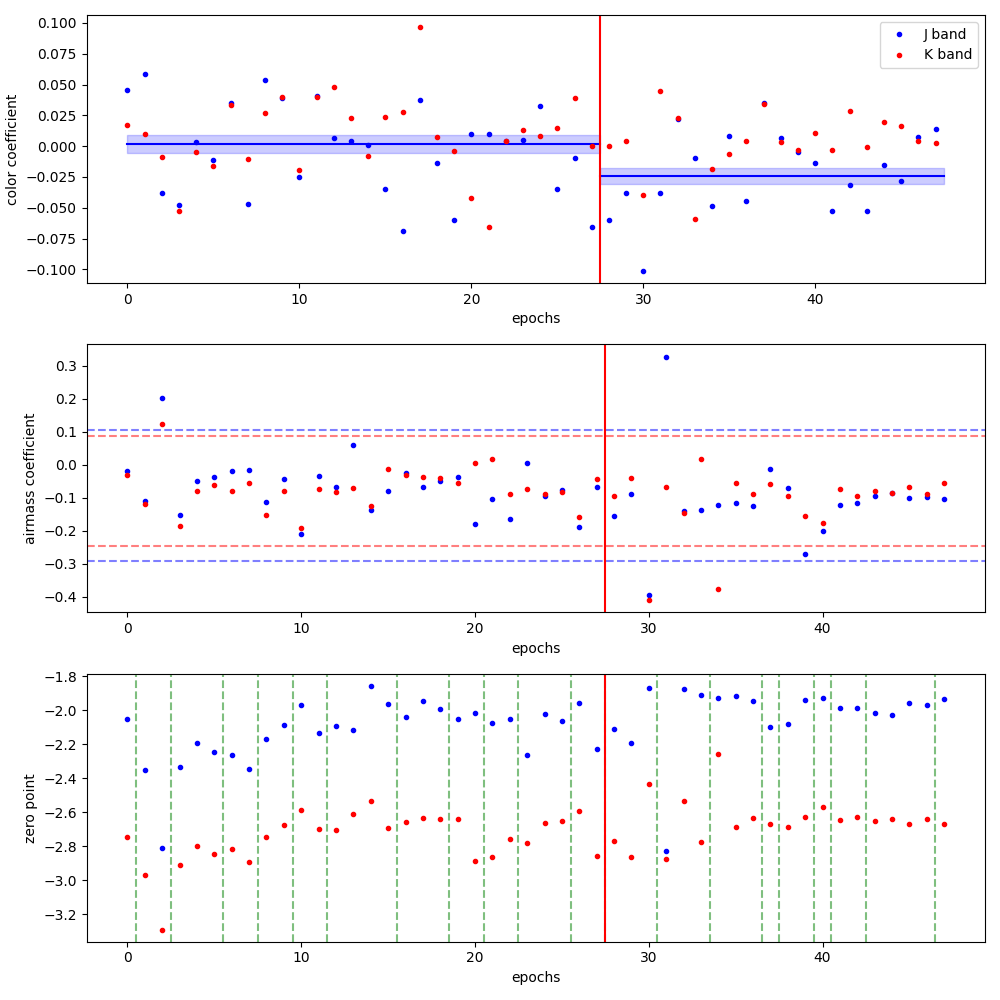}
\caption{Equation \ref{eq:transformation} color-term coefficient (upper panel), airmass coefficient (middle panel), and zero-point (lower panel) values obtained for all nights (observing epochs) using the free-fit approach. Blue and red points represent values for the J- and K-band, respectively. The blue horizontal lines in the upper panel mark the average values of the J-band color-term coefficient with their corresponding uncertainty (blue shaded area) for the periods up to and including 12 December 2013, and starting from 8 December 2014. Notably, the difference in the mean coefficient value reaches a significance level of $3\sigma$. The dotted blue and red lines in the middle panel mark the $3\sigma$ range for the J- and K-bands, respectively. The green dashed vertical lines in the lower panel indicate epochs where consecutive observations were separated by more than one week. The red vertical line between epochs 27 and 28 marks the transition corresponding to the change occurring between 12 December 2013 and 8 December 2014, as discussed in the text. The correspondence between observing epochs and calendar dates is given in Table \ref{tab:coefficients}.}
\label{fig:coefficients_free}
\end{figure}

Standard stars observations analyzed in this work were originally used to transform J- and K-band instrumental magnitudes (lower case: $j$ and $k$, respectively) of other objects onto the UKIRT standard system (upper case: $J$ and $K$, respectively). Transformations were carried out following the relations \eqref{eq:transformation}.

\begin{equation}
\label{eq:transformation}
\begin{aligned}
J &= j + c_{J} (j-k) + k_J \chi + z_J \\
K &= k + c_{K} (j-k) + k_K \chi + z_K ,
\end{aligned}
\end{equation}
where $\chi$ is the airmass at which the observations were executed and $j-k$ is the instrumental color of the star. A set of color-term coefficients ($c_{J}$, $c_{K}$), airmass coefficient ($k_J$, $k_K$) and zero points ($z_J$, $z_K$) were calculated each night using the least-squere method, adopting $J$ and $K$ from the \citet{Leggett2006} catalog. The values of the coefficients calculated for each night are presented in Figure \ref{fig:coefficients_free}.

Unfortunately, the uncertainty of the derived coefficients can be large, especially when an insufficient number of standard stars was observed on a given night. In fact, one could argue that the airmass coefficient should not vary more than 10\% from night to night under photometric conditions \citep{Burki1995}, while the color coefficient should change only if significant modifications were made to the instrumental system. Based on Figure \ref{fig:coefficients_free}, we suspect that such a change may have occurred between 12 December 2013 and 8 December 2014.

To verify this, we divided the entire observational period into two groups and calculated the mean value of the coefficient along with its uncertainty. We tested different separation dates, ensuring that no subsequent observations occurred within one month of the division date. Indeed, splitting the epochs into two groups, period up to 12 December 2013, and second period, starting from 8 December 2014, resulted in the largest difference in the mean coefficient value, reaching a level of 3 sigma. This division also minimized the scatter within both separated groups for both filters. We note that we do not observe similar effect for airmass coefficient, neither there was a need for more than two groups for color coefficient. 

To minimize large variations in calibrated magnitudes caused by large fluctuations in the color and airmass coefficients, we fit a single airmass coefficient for the entire 10-year observational period and allow only two values of the color coefficient, separated into two periods (up to and including 12 December 2013, and starting from 8 December 2014), independently for both bands.

We solve these equations algebraically by constructing the design matrix of the form:

\[
\begin{bmatrix}
(j-k)_{1,1} & 0 & \chi_{1,1} & 1 & 0 & \dots & 0 \\
(j-k)_{2,1} & 0 & \chi_{2,1} & 1 & 0 & \dots & 0 \\
\vdots & \vdots & \vdots & \vdots & \vdots & \ddots & \vdots \\
(j-k)_{m,1} & 0 & \chi_{m,1} & 1 & 0 & \dots & 0 \\
0 & (j-k)_{1,2} & \chi_{1,2} & 0 & 1 & \dots & 0 \\
0 & (j-k)_{2,2} & \chi_{2,2} & 0 & 1 & \dots & 0 \\
\vdots & \vdots & \vdots & \vdots & \vdots & \ddots & \vdots \\
0 & (j-k)_{m,n} & \chi_{m,n} & 0 & 0 & \dots & 1
\end{bmatrix}
\]

where $m$ denotes the ordinal number of the standard star observed on a given night and
$n$ represents the ordinal number of the night.

\begin{figure}[ht!]
\plotone{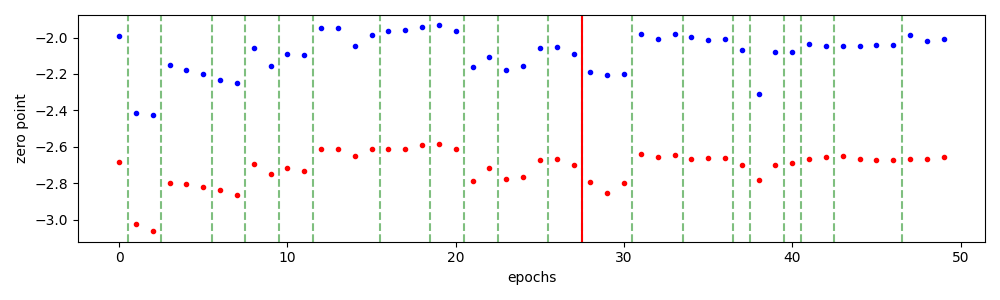}
\caption{Equation \ref{eq:transformation} zero-point values obtained for all nights (observing epochs) using the general least-squares fitting using a single value of the airmass coefficient for the entire 49-epoch observational period, and two values of the color coefficient separated into two periods: up to and including 12 December 2013, and starting from 8 December 2014, for the J- and K- bands (blue and red points, respectively). The green dashed vertical lines indicate epochs where consecutive observations were separated by more than one week. The red vertical line between epochs 27 and 28 marks the transition corresponding to the change of the color-term coefficient occurring between 12 December 2013 and 8 December 2014. The correspondence between observing epochs and calendar dates is given in Table \ref{tab:coefficients}. It can be noted that zero-point variations are much smaller compared to the free-fit results presented in Figure \ref{fig:coefficients_free}.}
\label{fig:coefficients}
\end{figure}

Figure \ref{fig:coefficients} presents the zero-point values calculated using this method, and Table \ref{tab:coefficients} reports their numerical values.

\begin{deluxetable*}{c|c|cc|cc}[ht!]
\tabletypesize{\scriptsize}
\tablewidth{0pt} 
\tablecaption{Equation \ref{eq:transformation} coefficients}
\tablehead{
\colhead{epoch} & \colhead{date} & \colhead{J-band zero point ($z_J$)} & \colhead{J rms} & 
\colhead{K-band zero point ($k_K$)} & \colhead{K rms} 
} 
\startdata 
\multicolumn{2}{c}{(epochs: 0-49 airmass coefficients:} & \multicolumn{2}{c}{$k_J$ = -0.0774 $\pm$ 0.0048} & \multicolumn{2}{c}{$k_K$ = -0.0786 $\pm$ 0.0057) } \\
\hline
\multicolumn{2}{c}{(epochs: 0-27 color coefficients:} & \multicolumn{2}{c}{$c_J$ = -0.0157 $\pm$ 0.0052} & \multicolumn{2}{c}{$c_K$ = 0.0011 $\pm$ 0.0063) } \\
 \hline
0 & 2008-12-13 & -1.990 $\pm$ 0.015 & 0.025 & -2.682 $\pm$ 0.019 & 0.011 \\
1 & 2009-11-5 & -2.414 $\pm$ 0.013 & 0.017 & -3.024 $\pm$ 0.016 & 0.013 \\
2 & 2009-11-7 & -2.424 $\pm$ 0.016 & 0.049 & -3.065 $\pm$ 0.020 & 0.063 \\
3 & 2009-12-2 & -2.152 $\pm$ 0.013 & 0.015 & -2.798 $\pm$ 0.016 & 0.023 \\
4 & 2009-12-3 & -2.177 $\pm$ 0.011 & 0.019 & -2.805 $\pm$ 0.014 & 0.020 \\
5 & 2009-12-4 & -2.199 $\pm$ 0.011 & 0.032 & -2.823 $\pm$ 0.013 & 0.026 \\
6 & 2009-12-26 & -2.231 $\pm$ 0.014 & 0.022 & -2.838 $\pm$ 0.017 & 0.016 \\
7 & 2009-12-28 & -2.251 $\pm$ 0.013 & 0.017 & -2.867 $\pm$ 0.016 & 0.045 \\
8 & 2011-12-30 & -2.059 $\pm$ 0.015 & 0.015 & -2.694 $\pm$ 0.018 & 0.017 \\
9 & 2011-12-31 & -2.157 $\pm$ 0.014 & 0.042 & -2.747 $\pm$ 0.017 & 0.031 \\
10 & 2012-1-6 & -2.092 $\pm$ 0.013 & 0.029 & -2.714 $\pm$ 0.016 & 0.022 \\
11 & 2012-1-7 & -2.098 $\pm$ 0.013 & 0.032 & -2.734 $\pm$ 0.016 & 0.027 \\
12 & 2012-10-10 & -1.949 $\pm$ 0.011 & 0.023 & -2.609 $\pm$ 0.013 & 0.031 \\
13 & 2012-10-11 & -1.947 $\pm$ 0.011 & 0.022 & -2.610 $\pm$ 0.014 & 0.035 \\
14 & 2012-10-12 & -2.045 $\pm$ 0.016 & 0.087 & -2.650 $\pm$ 0.019 & 0.058 \\
15 & 2012-10-13 & -1.986 $\pm$ 0.028 & 0.000 & -2.614 $\pm$ 0.035 & 0.000 \\
16 & 2012-11-1 & -1.964 $\pm$ 0.012 & 0.016 & -2.611 $\pm$ 0.015 & 0.032 \\
17 & 2012-11-2 & -1.960 $\pm$ 0.012 & 0.022 & -2.614 $\pm$ 0.014 & 0.028 \\
18 & 2012-11-3 & -1.940 $\pm$ 0.012 & 0.018 & -2.593 $\pm$ 0.014 & 0.012 \\
19 & 2012-11-15 & -1.933 $\pm$ 0.010 & 0.025 & -2.587 $\pm$ 0.013 & 0.028 \\
20 & 2012-11-16 & -1.965 $\pm$ 0.011 & 0.022 & -2.613 $\pm$ 0.013 & 0.028 \\
21 & 2013-8-24 & -2.164 $\pm$ 0.015 & 0.016 & -2.790 $\pm$ 0.019 & 0.042 \\
22 & 2013-8-25 & -2.108 $\pm$ 0.017 & 0.016 & -2.715 $\pm$ 0.021 & 0.019 \\
23 & 2013-11-26 & -2.180 $\pm$ 0.011 & 0.017 & -2.778 $\pm$ 0.013 & 0.021 \\
24 & 2013-11-27 & -2.159 $\pm$ 0.012 & 0.016 & -2.768 $\pm$ 0.014 & 0.010 \\
25 & 2013-11-28 & -2.058 $\pm$ 0.011 & 0.033 & -2.675 $\pm$ 0.013 & 0.016 \\
26 & 2013-12-11 & -2.049 $\pm$ 0.011 & 0.017 & -2.665 $\pm$ 0.013 & 0.018 \\
27 & 2013-12-12 & -2.091 $\pm$ 0.011 & 0.027 & -2.700 $\pm$ 0.013 & 0.018 \\
\hline
\multicolumn{2}{c}{(epochs: 28-49 color coefficients:} & \multicolumn{2}{c}{$c_J$ = -0.0362 $\pm$ 0.0034} & \multicolumn{2}{c}{$c_K$ = -0.0131 $\pm$ 0.0041) } \\
\hline
28 & 2014-12-8 & -2.1868 $\pm$ 0.0112 & 0.035 & -2.7915 $\pm$ 0.0136 & 0.034 \\
29 & 2014-12-9 & -2.2046 $\pm$ 0.0122 & 0.030 & -2.8518 $\pm$ 0.0151 & 0.116 \\
30 & 2014-12-10 & -2.2008 $\pm$ 0.0120 & 0.021 & -2.7969 $\pm$ 0.0145 & 0.017 \\
31 & 2015-9-26 & -1.9787 $\pm$ 0.0116 & 0.020 & -2.6394 $\pm$ 0.0139 & 0.014 \\
32 & 2015-9-27 & -2.0097 $\pm$ 0.0104 & 0.024 & -2.6559 $\pm$ 0.0127 & 0.027 \\
33 & 2015-9-28 & -1.9783 $\pm$ 0.0138 & 0.010 & -2.6463 $\pm$ 0.0169 & 0.067 \\
34 & 2015-12-19 & -1.9994 $\pm$ 0.0119 & 0.020 & -2.6671 $\pm$ 0.0145 & 0.023 \\
35 & 2015-12-20 & -2.0112 $\pm$ 0.0118 & 0.021 & -2.6621 $\pm$ 0.0143 & 0.017 \\
36 & 2015-12-21 & -2.0105 $\pm$ 0.0129 & 0.023 & -2.6595 $\pm$ 0.0162 & 0.016 \\
37 & 2016-6-10 & -2.0690 $\pm$ 0.0176 & 0.012 & -2.6975 $\pm$ 0.0215 & 0.002 \\
38 & 2016-6-26 & -2.3075 $\pm$ 0.0177 & 0.052 & -2.7810 $\pm$ 0.0222 & 0.040 \\
39 & 2016-6-27 & -2.0787 $\pm$ 0.0118 & 0.031 & -2.7013 $\pm$ 0.0142 & 0.014 \\
40 & 2017-9-7 & -2.0772 $\pm$ 0.0111 & 0.028 & -2.6866 $\pm$ 0.0135 & 0.030 \\
41 & 2017-9-21 & -2.0376 $\pm$ 0.0122 & 0.019 & -2.6656 $\pm$ 0.0149 & 0.020 \\
42 & 2017-9-22 & -2.0482 $\pm$ 0.0141 & 0.020 & -2.6552 $\pm$ 0.0172 & 0.014 \\
43 & 2018-11-18 & -2.0475 $\pm$ 0.0116 & 0.019 & -2.6530 $\pm$ 0.0141 & 0.024 \\
44 & 2018-11-19 & -2.0440 $\pm$ 0.0123 & 0.014 & -2.6679 $\pm$ 0.0150 & 0.029 \\
45 & 2018-11-20 & -2.0385 $\pm$ 0.0113 & 0.021 & -2.6731 $\pm$ 0.0138 & 0.031 \\
46 & 2018-11-21 & -2.0391 $\pm$ 0.0122 & 0.035 & -2.6707 $\pm$ 0.0153 & 0.021 \\
47 & 2018-12-26 & -1.9877 $\pm$ 0.0108 & 0.013 & -2.6678 $\pm$ 0.0132 & 0.026 \\
48 & 2018-12-27 & -2.0164 $\pm$ 0.0104 & 0.022 & -2.6652 $\pm$ 0.0127 & 0.028 \\
49 & 2018-12-28 & -2.0098 $\pm$ 0.0113 & 0.021 & -2.6571 $\pm$ 0.0137 & 0.024 
\enddata
\end{deluxetable*}
\label{tab:coefficients}

Finally, we applied Equation \ref{eq:transformation} to instrumental magnitudes to obtain standarized magnitudes for all stars in the fields, for each observing epoch.

\section{selection of secondary standards} \label{sec:selection}
Our goal was to prepare a catalog of selected secondary standards with the highest possible photometric quality while ensuring ease of use without accounting for any additional effects. 

To select stars with the best photometry, we used the RMS calculated in Section \ref{sec:phot} and the uncertainty of the mean value of the standardized magnitude from Section \ref{sec:standarization}. Every star in the final list met the following conditions:

\begin{itemize}
\item The standardized J- and K-band magnitudes are measured in at least five epochs.
\item The uncertainty of the average standardized magnitude is below 0.01 mag for both J- and K-band simultaneously. The uncertainty is calculated as $\sigma_{\bar{x}}$ $=$  $s$ / ${\sqrt{N}}$, where $s$ is the standard deviation and $N$ is the number of epochs. 
\item The RMS of the differential photometry across all epochs is below 0.03 mag for both J- and K-band simultaneously.
\item There is no excess of photometric noise in the J- and K-bands. This condition was applied using the same technique as described in Section \ref{sec:phot}: A star is excluded if its RMS value exceeds the corresponding value from the fitted relation of RMS versus the formal DAOPHOT error (Equation \ref{eq:err}) by more than 0.02 mag.
\end{itemize}

\begin{figure}[ht!]
\plotone{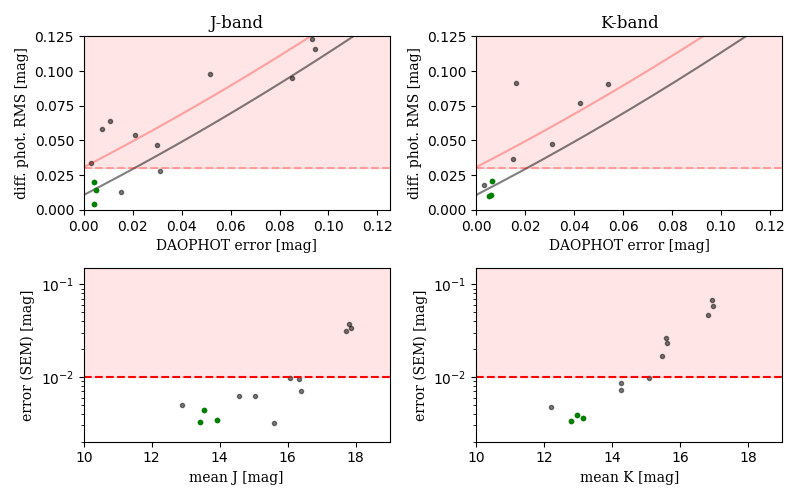}
\caption{Photometric quality selection criteria visualized for stars in the fields FS001. 
Upper panels shows RMS versus the average DAOPHOT error for all stars in the exemplary field for J- and K-band. Red dashed horizontal line indicates RMS value of 0.03 mag, and shaded red color indicates rejection area. Black lines represent the fitted models (eq. \ref{eq:err}) of RMS vs. DAOPHOT error relation used for excess noise estimation. Red solid line is shifted by 0.02 mag compared to the black line and indicates rejection condition. Lower panels show standard error of the mean (SEM) vs. mean value of the standardized J- and K-bands (left and right panels, respectively). Red dashed horizontal line indicates SEM value of 0.01 mag, and shaded red color indicates rejection area. Green points are stars that meet all the photometric conditions for secondary standards. 
\label{fig:selection}}
\end{figure}

We note that all rejected stars failed at least two of the four necessary conditions. Figure \ref{fig:selection} visualizes the photometric selection criteria for stars in the exemplary field FS001. 

With the preselected list of stars that meet the photometric conditions for all fields, we applied additional criteria:

\begin{itemize} 
\item Stars were rejected if there was a neighboring star closer than 6''. 
\item The star's parallax (\textsc{astroquery} GAIA \texttt{parallax}) must satisfy $\varpi / \sigma_\varpi > 1$,
where $\varpi$ is the parallax and $\sigma_\varpi$ is its uncertainty.
\item The proper motion of the star (\textsc{astroquery} GAIA \texttt{pmra} and \texttt{pmdec}) must not exceed 100 \text{mas/year}. 
\item No variability flag (\textsc{astroquery} GAIA \texttt{vari\_classifier\_result}) can be assigned to the star. 
\end{itemize}

The steps listed above involved querying Gaia DR3 data \citep{Gaia2023}.

\section{results and discussion} \label{sec:results}
\begin{figure*}[ht!]
\plotone{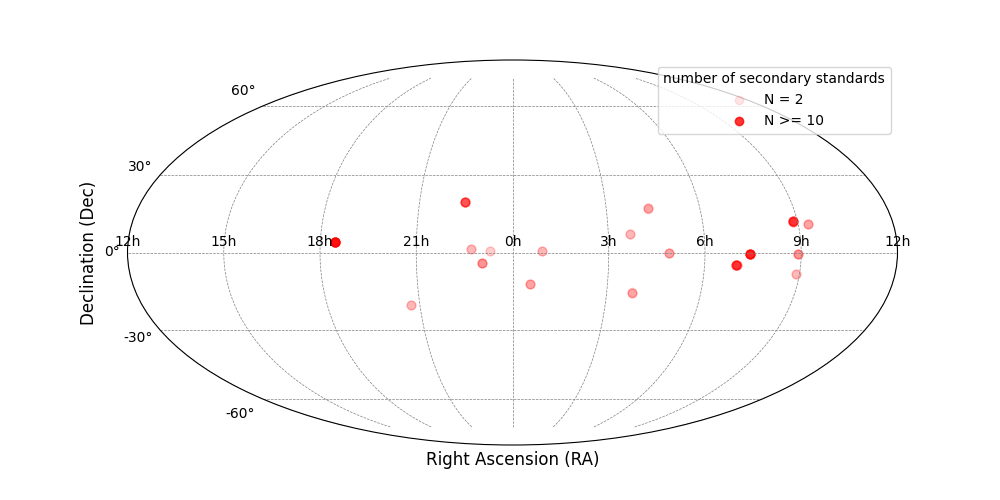}
\caption{Location of the 19 UKIRT faint standard fields with secondary standards defined in this work. The sky map in RA/Dec Coordinates is in the Mollweide Projection. The intensity of the color of the points indicates the number of secondary standards in the field. 
\label{fig:skymap}}
\end{figure*}

Based on the criteria described in the previous section, we prepared a catalog of 128 secondary standards in 19 UKIRT faint standard fields. Figure \ref{fig:skymap} shows the location of those fields in the sky. 

In Appendix \ref{sec:sec_std} we provide detailed information for each field, with the Finding Chart with marked positions of all secondary standards, color-magnitude diagrams of all secondary standards presented in this work, secondary standards for particular field (red points) and primary standard from the list \citet{Leggett2006} for the J- and K-bands. Finally, for each field, we provide a table with the secondary standard assigned name, ICRS RA/Dec coordinates for epoch 2000, J- and K-band magnitudes with corresponding uncertainty, number of observational epochs and Gaia IDs. The listed $J$ and $K$ magnitudes represent the mean values calculated across all available epochs, while the associated errors correspond to the standard error of the mean.

All of these products are also available on the Araucaria Project website (\url{araucaria.camk.edu.pl}) in additional formats. 

\subsection{Re-standarization of the primary standards\label{subsec:restandarization}}

\begin{figure}[ht!]
\plotone{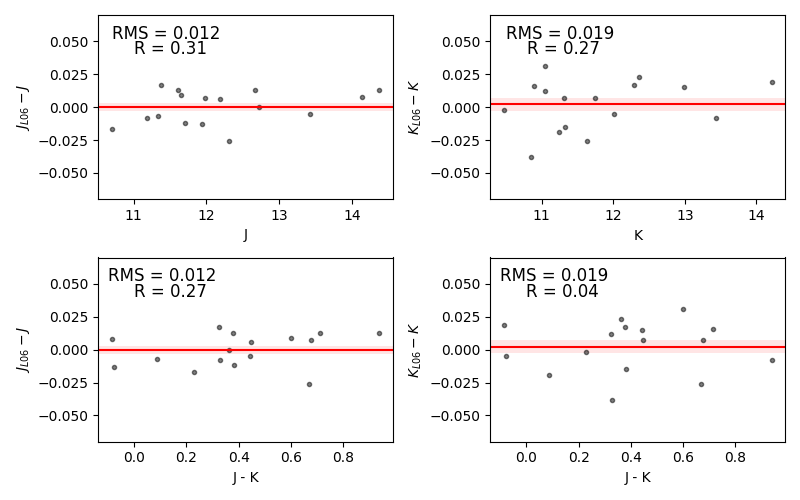}
\caption{The difference between the catalog values of the primary standards ($J_{L06}$, $K_{L06}$) and the mean values calculated in this work ($J$ and $K$). The catalog values are from \citet{Leggett2006}. The differences are plotted against magnitudes (upper panels) and $J-K$ color (lower panels). The red horizontal line indicates the mean value of all points, and the red shaded area represents the error of the mean. The Pearson correlation coefficient (R) and standard deviation (RMS) of the residuals are also reported.
\label{fig:l06}}
\end{figure}

The J- and K-band magnitudes presented in this paper were transformed into the MKO system using Equation \ref{eq:transformation} and using coefficients from Table \ref{tab:coefficients}. A
comparison between the transformed magnitudes of the primary standards and the catalog values provided by \citep{Leggett2006} serves as a basic consistency check for the procedure (Figure \ref{fig:l06}). The average difference across all points is consistent with zero within the calculated errors of the mean. The small values of the Pearson correlation coefficient (R) suggest that there is
no significant relation between the residuals and color or brightness. The average errors for the catalog data ($\sigma_{L06,J}$ = 0.009, $\sigma_{L06,K}$ = 0.010) and the average errors of re-standardized magnitudes presented in this study ($\sigma_{L06,J}$ = 0.006, $\sigma_{L06,K}$ = 0.005) are consistent with the observed scatter in the J-band. However, for the K-band, the scatter is approximately twice as large, which may indicate an underestimation of the errors provided by \citep{Leggett2006}, calculated in this work, or could suggest a difference between the MKO and NTT/SOFI photometric systems.

\subsection{Comparison with 2MASS \label{subsec:2mass}}
\begin{figure}[ht!]
\plotone{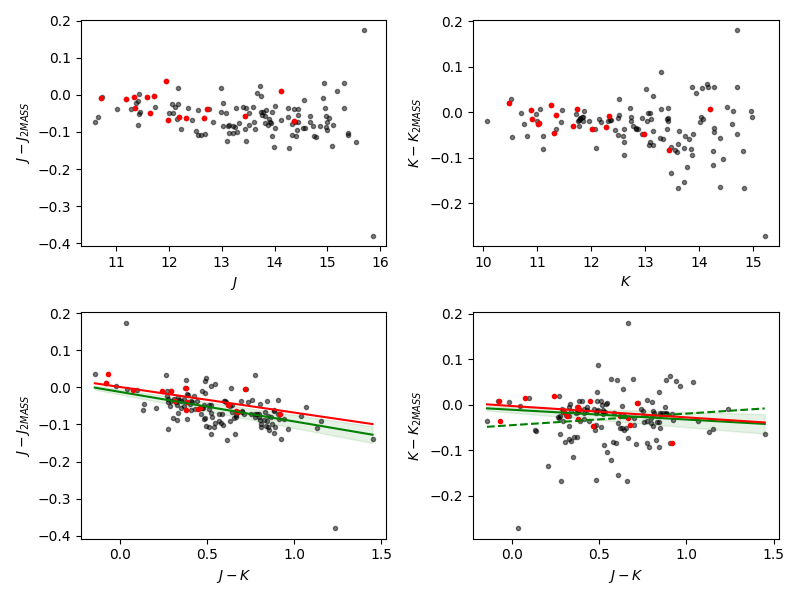}
\caption{The difference between the magnitudes obtained in this work ($J$ and $K$) and the 2MASS catalog values ($J_\mathrm{2MASS}$ and $K_\mathrm{2MASS}$). The upper panels show the magnitude differences as a function of magnitude, while the lower panels present them as a function of $J-K$ color. Red points represent primary standards, and black points correspond to secondary standards. The red solid lines in the lower panels indicate the color-based transformation between the MKO and 2MASS systems, as provided by \citet{Leggett2006}. The green solid lines represent the transformations derived in this study. The K-band transformation was obtained by fitting a linear relation to objects brighter than 13.5 mag in the K-band. The green dashed line illustrates the transformation when all data points are included.
\label{fig:2mass}}
\end{figure}

In this subsection, we compare our data ($J$ and $K$) with the magnitudes of the 2MASS catalog ($J_{2MASS}$ and $K_{2MASS}$). Figure \ref{fig:2mass} presents the magnitude differences as a function of $J-K$ color and magnitude. In both bands, a systematic shift in magnitudes is observed, with a small but noticeable color dependence in the J-band. While the spread of differences in the J-band remains uniform across the entire magnitude range, in the K-band, it increases for magnitudes fainter than 13.5.

\citet{Leggett2006} provide coefficients for the color-based transformation between the MKO and 2MASS systems in their Table 4. Using a least-squares method, we derived transformation coefficients based on our data. The slope and zero-point for the J-band are consistent with the values reported by \citet{Leggett2006} within the fitting uncertainties. However, for the K-band, the slope of the relation has the opposite sign when all data points are considered. Nevertheless, the slope remains statistically consistent with zero within the uncertainties of the fit. This is a consequence of the relatively large photometric scatter for fainter sources ($K > 13.5$), which limits the precision of the derived transformation coefficients. When limiting the data set to objects with a smaller scatter (K-band magnitudes brighter than 13.5), the resulting slope and zero-point agree with the values from \citet{Leggett2006}, and are given by:

\begin{eqnarray*}
\label{eq:tr2mass}
J - J_{\mathrm{2MASS}} &= (-0.080 \pm 0.011) \cdot (J - K) - (0.012 \pm 0.007), \\
K - K_{\mathrm{2MASS}} &= (-0.021 \pm 0.010) \cdot (J - K) - (0.011 \pm 0.006). 
\end{eqnarray*}

This agreement indicates that our photometric calibration is robust and that the SOFI/NTT system behaves consistently and can be reliably tied to 2MASS through the MKO framework.

\subsection{Deriving transformation coefficients with alternative approaches}

In Section \ref{sec:standarization}, we derived the transformation coefficients of the photometric system by allowing only a single airmass coefficient and two color coefficients per band across all epochs.

If, instead of this procedure, we allow these coefficients to be fitted individually for each night, the average magnitudes of the secondary standards remain virtually the same. However, the spread of magnitudes from night to night increases by 30\%. As a result, the estimated uncertainty of the calculated average magnitudes is larger.

Additionally, we explored other coefficients combining procedures, such as grouping the airmass (and color) coefficients by month or by observing run. In all cases of combining multiple epochs, the resulting spread in magnitudes was significantly smaller than in the free-fit procedure. For consistency and clarity, we ultimately decided to adopt the procedure with one airmass and two color coefficients per band. 

\subsection{Calibration of secondary standards relative to the primary standard}

In this work, we chose to calculate the magnitudes of all secondary standards separately, using calibration coefficients derived for each individual night.

An alternative approach would be to calibrate the brightness of secondary standards based on the brightness of the primary standard in a given field, accounting for instrumental magnitude differences, and applying a color correction. Although this method could improve statistical accuracy for some limited number of objects by roughly 0.001 mag, it would also introduce a systematic error for all stars in the field, comparable to the statistical uncertainty of the primary standard's brightness statistical uncertainty. 

\subsection{Consistency across RA coverage}
The two extreme ends of our RA coverage were not observed on the same night, preventing a direct measurement of potential zero-point differences between them. 
The use of overlapping standards observed on different nights provides a robust connection of zero-points across the entire RA span. 
Thus, even though the largest RA gap in our sample is approximately 9 hours, the use of overlapping standards maintains a consistent photometric system across the field.

\section{summary and conclusions} \label{sec:summary}
We presented a catalog of 128 secondary standard stars located in 19 UKIRT/MKO faint standard fields, based on 10 years of Araucaria Project observations using the NTT telescope equipped with the SOFI NIR camera. The average J- and K-band magnitudes of these stars are calibrated to the MKO photometric system of \citet{Leggett2006}. The magnitudes range from 10 to 15.8, with medians of $\tilde{J}$ = 13.5 and $\tilde{K}$ = 13. The uncertainty in the brightness measurements is less than 0.01 mag for all stars. The $J-K$ colors of the secondary standards range from -0.07 to 1.4, with a median value of 0.53 mag. The number of newly defined secondary standards per field varies from 1 to 22, with fields FS121, FS035, and FS014 containing more than 10 stars each. Our results suggest that using these fields for standardization can improve the precision and accuracy of photometric calibrations without incurring additional observational-time costs.

\begin{acknowledgments}
The research leading to these results has received funding from the European Research Council (ERC) under the European Union’s Horizon 2020 research and innovation program (grant agreement No. 951549). We acknowledge the contribution of present and former collaborators and members of the Araucaria Project, whose collective effort made this database possible. Based on observations collected at the European Southern Observatory under ESO programme IDs:
190.D-0237(B,D), 095.D-0424(B), 092.D-0295(B), 090.D-0409(B), 
084.D-0591(E,B), 094.D-0056(B), 099.D-0307(A), 0102.D-0590(B), 
084.D-0640(B), 097.D-0151(A), 088.D-0447(B), 088.D-0401(B), 
0102.D-0469(B), 096.D-0170(B), 092.D-0349(A), 082.D-0513(A).
\end{acknowledgments}

%

\vspace{5mm}
\facilities{NTT}


\software{astropy \citep{astropy2022},  
        astroquery, 
          DAOPHOT,
          IRAF
          }


\bibliography{references}{}
\bibliographystyle{aasjournal}

\appendix

\section{Observation log and individual secondary standards fields} 
\label{app:obs_log}

\label{sec:sec_std}

\begin{deluxetable*}{l|c|c|c|c|c|c|c|c|c|c|c|c|c|c|c|c|c|c|c|}
\tablecaption{Observation log of UKIRT standard stars. "X" indicates that the respective standard star was observed on that night.
}
\tablewidth{700pt}
\tabletypesize{\scriptsize}
\tablehead{
\colhead{Date} & \colhead{\rotatebox{270}{FS001 }} & \colhead{\rotatebox{270}{FS002 }} & \colhead{\rotatebox{270}{FS011 }} & \colhead{\rotatebox{270}{FS014 }} & \colhead{\rotatebox{270}{FS015 }} & \colhead{\rotatebox{270}{FS017 }} & \colhead{\rotatebox{270}{FS018 }} & \colhead{\rotatebox{270}{FS030 }} & \colhead{\rotatebox{270}{FS034 }} & \colhead{\rotatebox{270}{FS035 }} & \colhead{\rotatebox{270}{FS110 }} & \colhead{\rotatebox{270}{FS112 }} & \colhead{\rotatebox{270}{FS114 }} & \colhead{\rotatebox{270}{FS121 }} & \colhead{\rotatebox{270}{FS124 }} & \colhead{\rotatebox{270}{FS126 }} & \colhead{\rotatebox{270}{FS152 }} & \colhead{\rotatebox{270}{FS153 }} & \colhead{\rotatebox{270}{FS154 }} 
} 
\startdata
2008-12-13 & X &   &   &   &   &   &   & X &   &   & X &   & X &   &   &   &   &   &   \\
2009-11-05 &   &   &   &   &   &   &   & X & X &   &   &   &   & X &   &   &   &   & X \\
2009-11-06 & X &   &   &   &   &   &   &   & X &   &   &   &   &   &   &   &   &   & X \\
2009-11-07 & X &   &   &   &   &   &   &   & X &   &   &   &   &   &   &   &   &   & X \\
2009-12-02 &   &   &   &   & X &   &   &   & X &   & X &   & X &   &   &   &   &   &   \\
2009-12-03 & X &   &   & X & X &   &   &   & X &   &   &   & X &   &   &   &   &   &   \\
2009-12-04 & X &   &   &   & X & X &   &   & X &   & X &   &   & X &   &   &   &   &   \\
2009-12-26 &   & X &   & X & X &   &   &   &   &   &   &   &   &   &   &   &   &   &   \\
2009-12-28 &   &   &   & X & X &   &   &   &   &   &   &   &   &   &   &   &   &   &   \\
2011-12-30 &   &   & X &   &   &   & X &   &   &   &   & X & X & X & X &   &   &   &   \\
2011-12-31 &   &   &   &   & X &   & X &   &   &   & X & X & X & X & X &   &   &   &   \\
2012-01-06 &   &   &   &   &   & X &   &   &   &   &   &   &   & X & X & X &   &   &   \\
2012-01-07 &   &   &   &   &   &   & X &   &   &   & X & X & X &   &   & X &   &   &   \\
2012-10-10 &   & X &   &   &   &   &   &   &   & X & X & X & X &   &   &   & X &   & X \\
2012-10-11 & X & X &   &   &   &   &   &   &   & X &   & X & X &   &   &   & X &   & X \\
2012-10-12 &   &   &   &   &   &   &   &   &   &   &   &   &   &   &   &   & X &   &   \\
2012-10-13 &   &   &   &   &   &   &   &   &   & X &   &   &   &   &   &   &   &   &   \\
2012-11-01 & X &   &   &   &   & X & X &   &   &   &   &   &   & X & X &   & X & X & X \\
2012-11-02 & X & X &   &   &   &   &   &   &   &   & X & X &   &   & X & X & X & X & X \\
2012-11-03 & X &   &   &   &   &   &   &   &   &   & X &   &   & X &   & X & X & X &   \\
2012-11-15 & X & X & X & X &   &   &   &   & X &   &   & X &   &   &   & X & X & X &   \\
2012-11-16 & X &   &   & X &   &   &   &   & X &   &   & X & X &   &   &   & X &   & X \\
2013-08-24 &   &   &   &   &   &   &   &   &   &   &   &   & X &   &   &   &   &   &   \\
2013-08-25 &   &   &   &   &   &   &   &   & X & X &   &   &   &   &   &   &   &   &   \\
2013-11-26 & X &   &   & X & X & X &   &   &   &   & X &   & X & X &   &   &   &   &   \\
2013-11-27 &   &   &   & X &   & X &   &   &   &   &   &   & X & X &   &   &   &   &   \\
2013-11-28 & X &   & X & X & X & X &   &   &   &   & X &   & X & X & X &   &   &   &   \\
2013-12-11 & X &   &   &   &   &   & X &   &   &   & X &   & X & X & X & X &   & X &   \\
2013-12-12 & X &   &   &   &   &   & X &   &   &   &   &   & X & X & X & X &   & X &   \\
2014-12-08 & X & X & X & X &   &   &   &   & X &   & X & X & X &   &   &   &   &   &   \\
2014-12-09 &   & X & X &   &   &   &   &   & X &   &   & X & X &   &   &   &   &   &   \\
2014-12-10 &   & X & X & X &   & X &   &   &   &   &   & X & X & X &   &   &   &   &   \\
2015-01-04 & X &   &   &   &   &   &   &   &   &   & X &   & X &   &   &   &   & X &   \\
2015-01-06 &   &   &   &   &   &   &   &   &   &   &   & X & X &   &   &   &   & X &   \\
2015-09-26 & X & X &   & X &   &   &   &   & X &   &   &   & X &   & X &   &   &   &   \\
2015-09-27 & X &   & X &   &   &   &   &   & X & X &   & X & X &   &   &   & X & X &   \\
2015-09-28 &   &   &   &   &   &   &   &   & X & X &   &   & X &   &   &   &   & X &   \\
2015-12-19 &   &   &   &   & X &   &   & X &   &   &   &   & X &   & X & X &   &   &   \\
2015-12-20 &   &   & X &   & X &   &   &   &   &   &   & X & X & X &   & X &   & X & X \\
2015-12-21 &   &   &   &   &   &   &   & X &   &   &   &   &   & X & X & X &   & X & X \\
2016-06-10 &   &   &   &   & X &   & X &   &   &   &   &   &   &   &   &   &   &   &   \\
2016-06-26 &   &   &   &   &   &   &   & X &   &   &   &   &   &   &   &   &   & X &   \\
2016-06-27 &   &   &   &   &   &   &   &   & X &   &   &   &   &   &   &   &   &   &   \\
2017-09-07 &   &   &   &   &   &   &   & X & X &   & X &   &   &   &   &   &   & X &   \\
2017-09-21 &   & X &   &   &   &   &   &   & X &   &   &   & X &   &   &   &   &   &   \\
2017-09-22 &   &   &   &   &   &   &   &   & X &   &   &   & X &   &   &   &   &   &   \\
2017-09-23 &   &   &   &   &   &   &   &   & X &   &   &   &   &   &   &   &   &   &   \\
2018-11-18 &   &   & X & X &   &   & X &   & X &   &   & X &   & X &   &   & X &   &   \\
2018-11-19 &   &   & X &   &   &   & X &   &   &   &   &   &   & X & X & X &   & X &   \\
2018-11-20 &   &   & X &   &   &   &   & X &   &   &   & X &   &   & X &   & X &   &   \\
2018-11-21 &   &   &   & X &   &   &   & X &   &   &   &   &   &   &   &   & X & X & X \\
2018-12-26 &   &   &   & X &   &   & X & X &   &   &   & X &   & X & X &   &   &   &   \\
2018-12-27 & X &   & X &   & X &   &   & X &   &   &   & X &   &   & X &   &   & X &   \\
2018-12-28 & X &   & X & X & X &   & X &   &   &   &   & X &   & X & X & X &   & X &   \\
\enddata
\end{deluxetable*}
\label{tab:obs_log}

\newpage
\onecolumngrid
In the appendix, we report  which standards were observed each night (Table \ref{tab:obs_log}) and provide detailed information for each of the 19 standard star fields. 

Finding charts are included, showing the positions of the primary standard (blue circle) and secondary standards (red circles) along with their names. Color-magnitude diagrams are presented for the J- and K-bands vs. the J-K color. In each diagram, all secondary standards defined in this work across all fields are shown as black dots, while the primary standard and the secondary standards for the given field are represented as blue and red dots, respectively. Finally, for each field, we provide a table containing the names, ICRS RA/Dec coordinates (epoch 2000), J- and K-band magnitudes standardized to the MKO system, along with their corresponding uncertainties, number of observational epochs and their GAIA IDs.
The primary standard is not shown for FS018 and FS124. In the case of FS018, the star is saturated in our observations, and no reliable photometry could be obtained due to observational limitations. The primary standard for FS124 was excluded from the final sample because of its high proper motion.

\newpage

\begin{figure*}[t]
\centering
\includegraphics[width=0.5\textwidth]{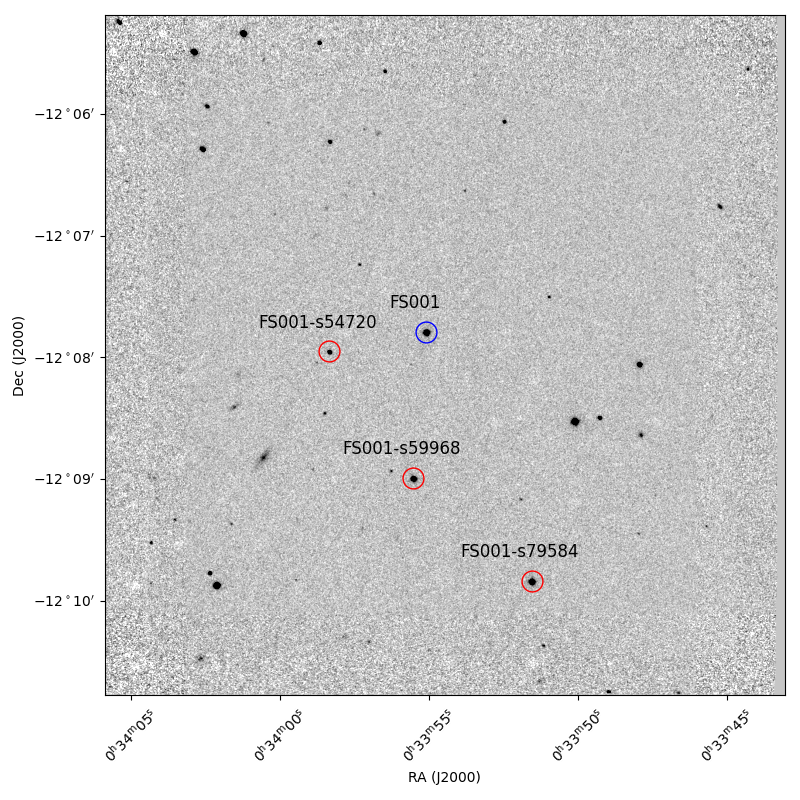} \\
\vspace{5pt}
\includegraphics[width=0.5\textwidth]{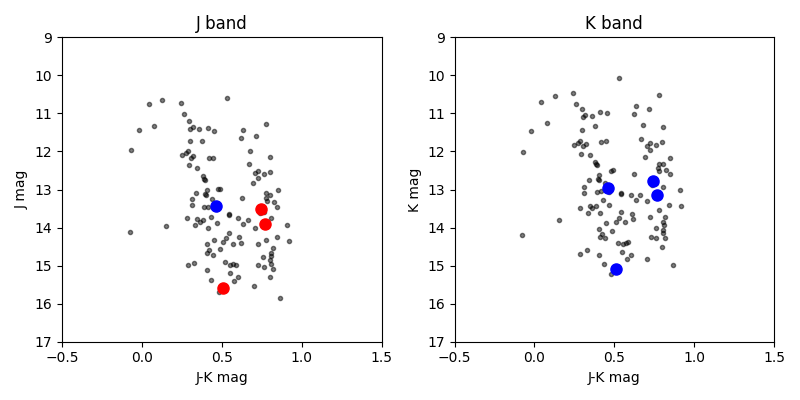}
\caption{FS001 field finding chart and color-magnitude diagrams}
\end{figure*}

\begin{deluxetable*}{ccc|ccc|ccc|c}[ht!]
\tabletypesize{\scriptsize}
\tablewidth{0pt} 
\tablecaption{FS001}
\tablehead{
\colhead{name} & \colhead{ra} & \colhead{dec} & 
\colhead{J} & \colhead{err J} & \colhead{epochs J} &
\colhead{K} & \colhead{err K} & \colhead{epochs K} &
\colhead{GAIA id}  
} 
\startdata 
FS001 & 0:33:54.46 & -12:07:58.78 & 13.432 & 0.003 & 17 & 12.969 & 0.004 & 17 & 2375647158466154112 \\
FS001-s79584 & 0:33:51.05 & -12:10:03.71 & 13.524 & 0.004 & 14 & 12.779 & 0.003 & 14 & 2375643688132579584 \\
FS001-s59968 & 0:33:55.04 & -12:09:13.14 & 13.915 & 0.003 & 17 & 13.146 & 0.004 & 17 & 2375643821276259968 \\
FS001-s54720 & 0:33:57.86 & -12:08:10.88 & 15.594 & 0.003 & 17 & 15.085 & 0.010 & 17 & 2375644203528654720 \\
\enddata
\end{deluxetable*}

\newpage

\begin{figure*}[t]
\centering
\includegraphics[width=0.5\textwidth]{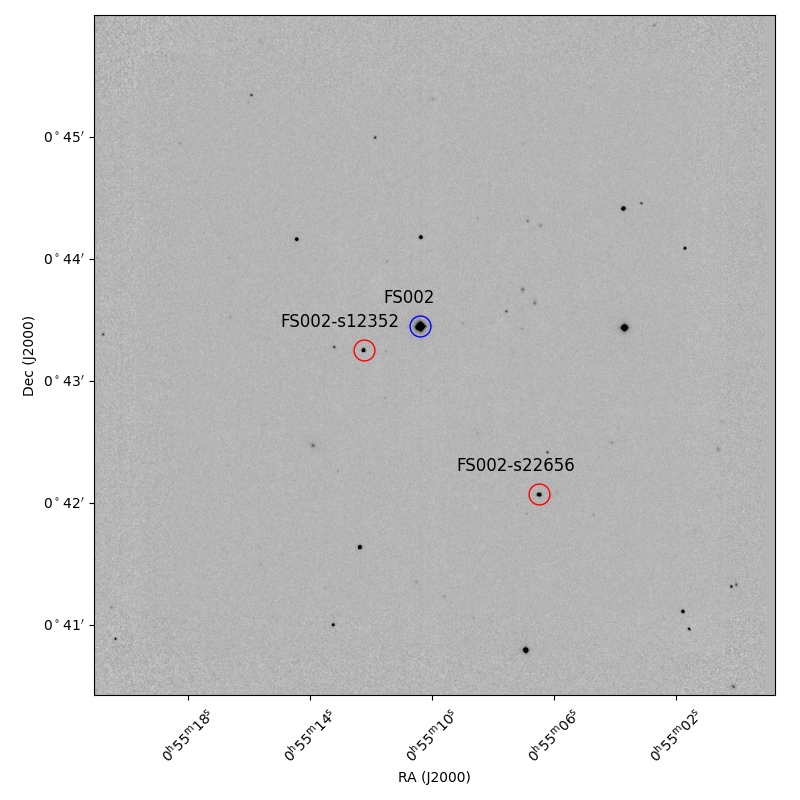} \\
\vspace{5pt}
\includegraphics[width=0.5\textwidth]{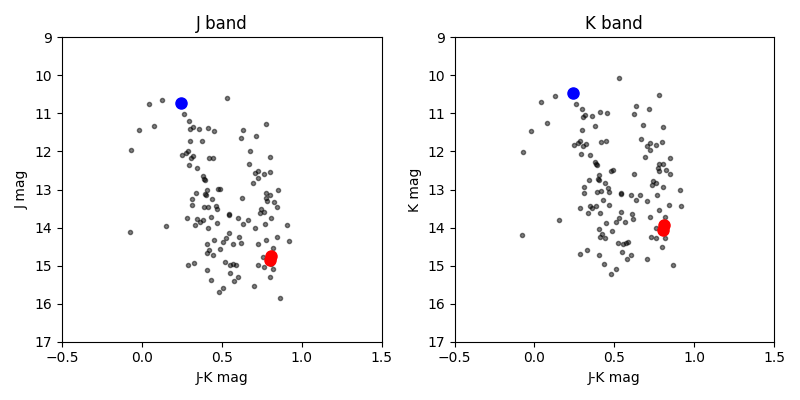}
\caption{FS002 field finding chart and color-magnitude diagrams}
\end{figure*}

\begin{deluxetable*}{ccc|ccc|ccc|c}[ht!]
\tabletypesize{\scriptsize}
\tablewidth{0pt} 
\tablecaption{FS002}
\tablehead{
\colhead{name} & \colhead{ra} & \colhead{dec} & 
\colhead{J} & \colhead{err J} & \colhead{epochs J} &
\colhead{K} & \colhead{err K} & \colhead{epochs K} &
\colhead{GAIA id}  
} 
\startdata 
FS002 & 0:55:09.91 & 0:43:12.92 & 10.716 & 0.003 & 7 & 10.472 & 0.004 & 7 & 2537314812728975744 \\
FS002-s22656 & 0:55:06.00 & 0:41:50.29 & 14.746 & 0.006 & 7 & 13.936 & 0.006 & 7 & 2537314675290022656 \\
FS002-s12352 & 0:55:11.75 & 0:43:01.25 & 14.862 & 0.006 & 7 & 14.059 & 0.005 & 7 & 2537314808433812352 \\
\enddata
\end{deluxetable*}

\newpage

\begin{figure*}[t]
\centering
\includegraphics[width=0.5\textwidth]{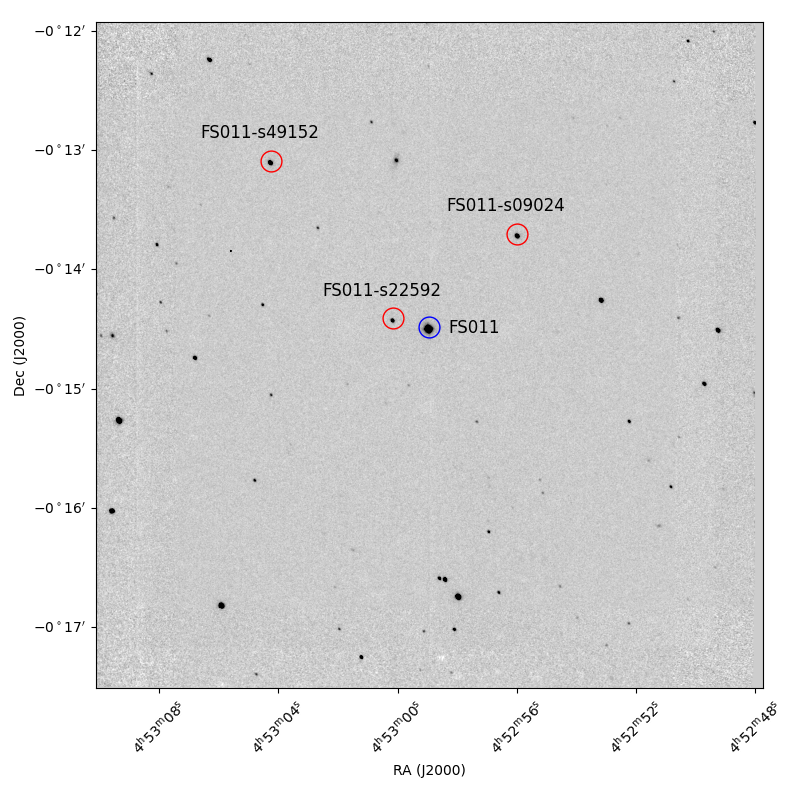} \\
\vspace{5pt}
\includegraphics[width=0.5\textwidth]{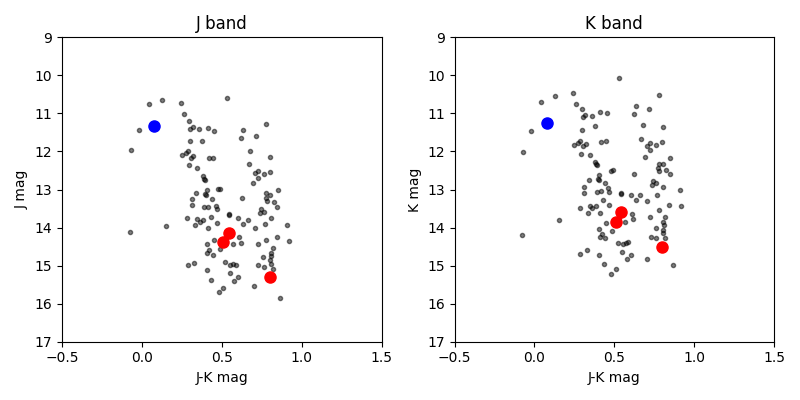}
\caption{FS011 field finding chart and color-magnitude diagrams}
\end{figure*}

\begin{deluxetable*}{ccc|ccc|ccc|c}[ht!]
\tabletypesize{\scriptsize}
\tablewidth{0pt} 
\tablecaption{FS011}
\tablehead{
\colhead{name} & \colhead{ra} & \colhead{dec} & 
\colhead{J} & \colhead{err J} & \colhead{epochs J} &
\colhead{K} & \colhead{err K} & \colhead{epochs K} &
\colhead{GAIA id}  
} 
\startdata 
FS011 & 4:52:58.86 & -0:14:41.17 & 11.336 & 0.005 & 10 & 11.260 & 0.004 & 10 & 3226810514329499648 \\
FS011-s09024 & 4:52:55.88 & -0:13:54.39 & 14.130 & 0.010 & 10 & 13.585 & 0.010 & 10 & 3226810720487809024 \\
FS011-s49152 & 4:53:04.16 & -0:13:17.76 & 14.373 & 0.008 & 9 & 13.865 & 0.006 & 9 & 3226810651768449152 \\
FS011-s22592 & 4:53:00.06 & -0:14:36.98 & 15.309 & 0.009 & 10 & 14.510 & 0.009 & 10 & 3226810510033822592 \\
\enddata
\end{deluxetable*}

\newpage

\begin{figure*}[t]
\centering
\includegraphics[width=0.45\textwidth]{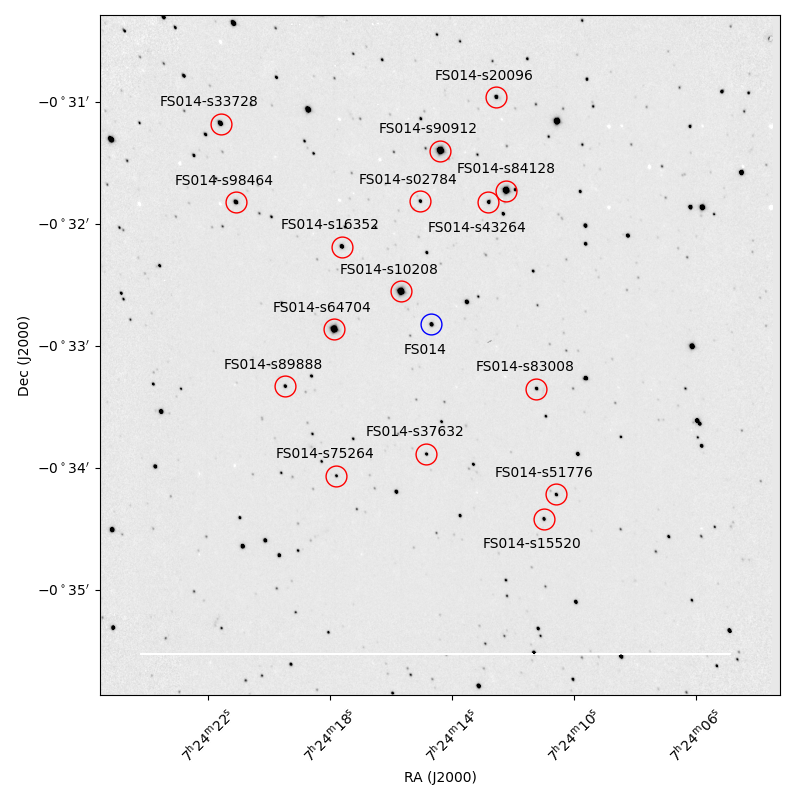} \\
\vspace{5pt}
\includegraphics[width=0.45\textwidth]{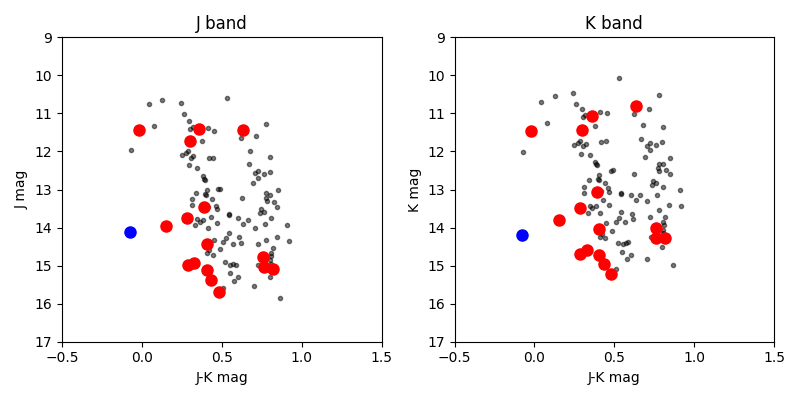}
\caption{FS014 field finding chart and color-magnitude diagrams}
\end{figure*}

\begin{deluxetable*}{ccc|ccc|ccc|c}[ht!]
\tabletypesize{\scriptsize}
\tablewidth{0pt} 
\tablecaption{FS014}
\tablehead{
\colhead{name} & \colhead{ra} & \colhead{dec} & 
\colhead{J} & \colhead{err J} & \colhead{epochs J} &
\colhead{K} & \colhead{err K} & \colhead{epochs K} &
\colhead{GAIA id}  
} 
\startdata 
FS014-s90912 & 7:24:14.08 & -0:31:38.68 & 11.442 & 0.007 & 13 & 10.808 & 0.007 & 13 & 3110405355740790912 \\
FS014-s10208 & 7:24:15.38 & -0:32:47.84 & 11.443 & 0.007 & 13 & 11.463 & 0.005 & 13 & 3110405183942110208 \\
FS014-s64704 & 7:24:17.57 & -0:33:06.18 & 11.415 & 0.006 & 13 & 11.055 & 0.005 & 13 & 3110404428027864704 \\
FS014-s84128 & 7:24:11.93 & -0:31:58.12 & 11.731 & 0.009 & 13 & 11.430 & 0.006 & 13 & 3110405252661584128 \\
FS014-s33728 & 7:24:21.29 & -0:31:25.28 & 13.462 & 0.007 & 13 & 13.072 & 0.006 & 13 & 3110405321381033728 \\
FS014-s98464 & 7:24:20.78 & -0:32:03.99 & 13.760 & 0.007 & 13 & 13.475 & 0.007 & 13 & 3110404565466798464 \\
FS014-s16352 & 7:24:17.32 & -0:32:25.83 & 13.957 & 0.007 & 13 & 13.805 & 0.008 & 13 & 3110405287021316352 \\
FS014 & 7:24:14.37 & -0:33:04.16 & 14.120 & 0.006 & 13 & 14.195 & 0.005 & 13 & 3110404393668131712 \\
FS014-s20096 & 7:24:12.25 & -0:31:12.40 & 14.441 & 0.008 & 13 & 14.035 & 0.010 & 13 & 3110405561899220096 \\
FS014-s43264 & 7:24:12.51 & -0:32:03.95 & 14.776 & 0.009 & 13 & 14.015 & 0.007 & 13 & 3110405248362843264 \\
FS014-s89888 & 7:24:19.19 & -0:33:34.35 & 15.036 & 0.006 & 13 & 14.272 & 0.006 & 13 & 3110404359308389888 \\
FS014-s83008 & 7:24:10.93 & -0:33:35.80 & 14.979 & 0.008 & 13 & 14.693 & 0.009 & 13 & 3110381681881083008 \\
FS014-s15520 & 7:24:10.70 & -0:34:39.83 & 14.929 & 0.007 & 13 & 14.602 & 0.008 & 13 & 3110380891607115520 \\
FS014-s02784 & 7:24:14.74 & -0:32:03.54 & 15.114 & 0.006 & 13 & 14.709 & 0.010 & 13 & 3110405183942102784 \\
FS014-s51776 & 7:24:10.30 & -0:34:28.00 & 15.094 & 0.006 & 13 & 14.276 & 0.010 & 13 & 3110380887308251776 \\
FS014-s37632 & 7:24:14.54 & -0:34:07.91 & 15.388 & 0.008 & 13 & 14.955 & 0.009 & 13 & 3110380925966837632 \\
FS014-s75264 & 7:24:17.49 & -0:34:18.50 & 15.696 & 0.010 & 13 & 15.216 & 0.007 & 13 & 3110404324950075264 \\
\enddata
\end{deluxetable*}


\newpage

\begin{figure*}[t]
\centering
\includegraphics[width=0.5\textwidth]{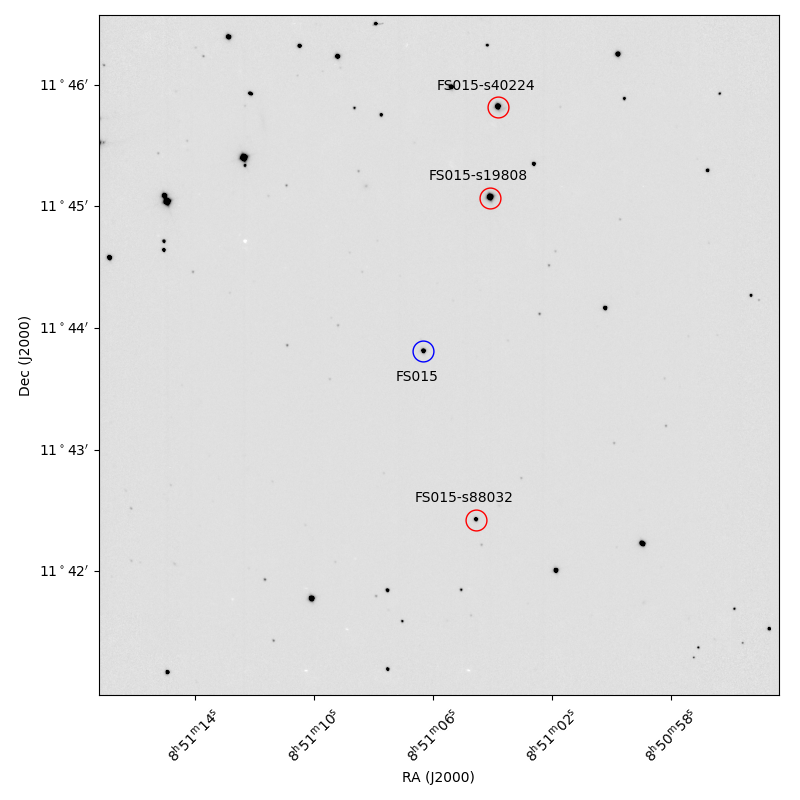} \\
\vspace{5pt}
\includegraphics[width=0.5\textwidth]{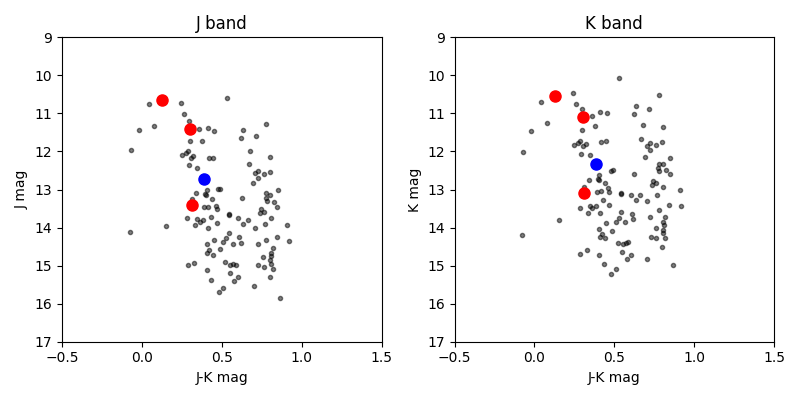}
\caption{FS015 field finding chart and color-magnitude diagrams}
\end{figure*}

\begin{deluxetable*}{ccc|ccc|ccc|c}[ht!]
\tabletypesize{\scriptsize}
\tablewidth{0pt} 
\tablecaption{FS015}
\tablehead{
\colhead{name} & \colhead{ra} & \colhead{dec} & 
\colhead{J} & \colhead{err J} & \colhead{epochs J} &
\colhead{K} & \colhead{err K} & \colhead{epochs K} &
\colhead{GAIA id}  
} 
\startdata 
FS015-s19808 & 8:51:03.51 & 11:45:02.82 & 10.658 & 0.008 & 6 & 10.530 & 0.005 & 6 & 604911135364519808 \\
FS015-s40224 & 8:51:03.26 & 11:45:47.41 & 11.409 & 0.008 & 5 & 11.105 & 0.003 & 5 & 604914468259140224 \\
FS015 & 8:51:05.76 & 11:43:46.97 & 12.722 & 0.010 & 6 & 12.336 & 0.008 & 6 & 604910860486613632 \\
FS015-s88032 & 8:51:03.99 & 11:42:23.95 & 13.408 & 0.010 & 6 & 13.097 & 0.007 & 6 & 604910654328188032 \\
\enddata
\end{deluxetable*}


\newpage

\begin{figure*}[t]
\centering
\includegraphics[width=0.45\textwidth]{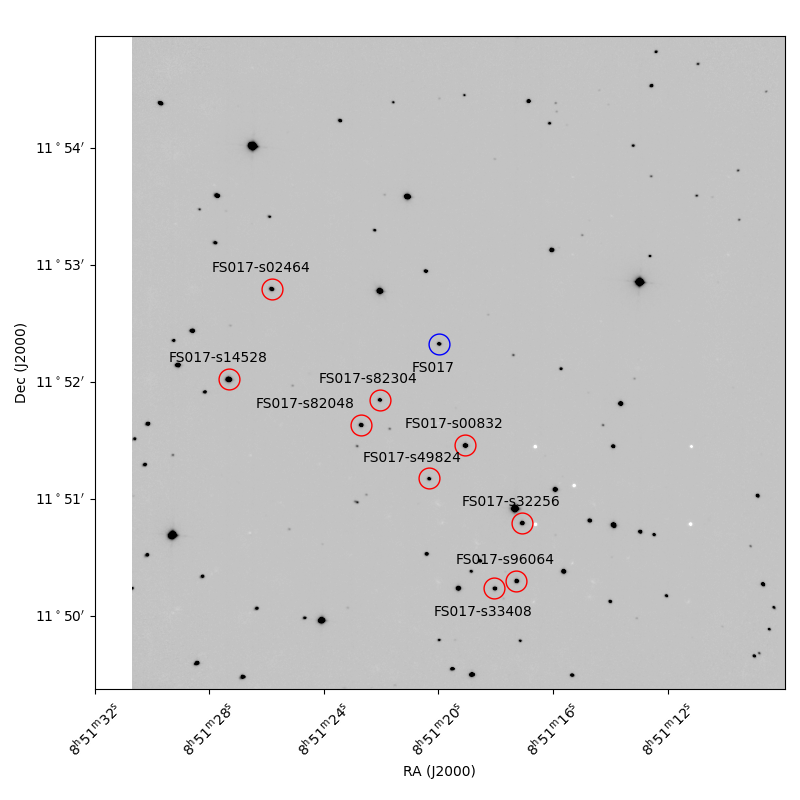} \\
\vspace{5pt}
\includegraphics[width=0.45\textwidth]{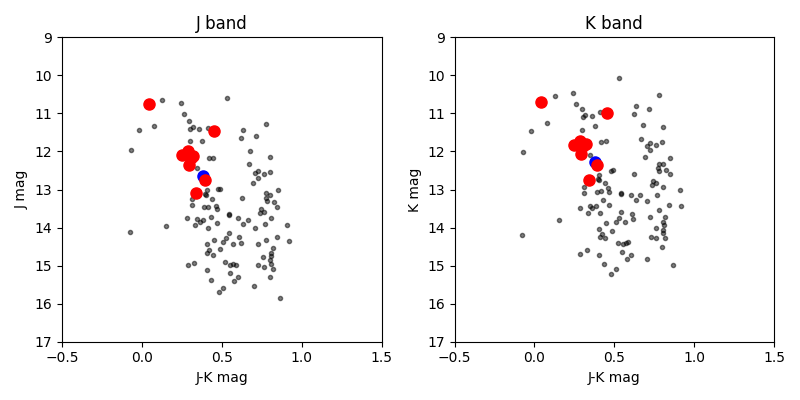}
\caption{FS017 field finding chart and color-magnitude diagrams}
\end{figure*}

\begin{deluxetable*}{ccc|ccc|ccc|c}[ht!]
\tabletypesize{\scriptsize}
\tablewidth{0pt} 
\tablecaption{FS017}
\tablehead{
\colhead{name} & \colhead{ra} & \colhead{dec} & 
\colhead{J} & \colhead{err J} & \colhead{epochs J} &
\colhead{K} & \colhead{err K} & \colhead{epochs K} &
\colhead{GAIA id}  
} 
\startdata 
FS017-s14528 & 8:51:27.01 & 11:51:52.58 & 10.742 & 0.009 & 7 & 10.699 & 0.005 & 7 & 604921202767814528 \\
FS017-s00832 & 8:51:18.77 & 11:51:18.71 & 11.455 & 0.008 & 7 & 11.001 & 0.003 & 7 & 604921129752600832 \\
FS017-s32256 & 8:51:16.79 & 11:50:39.02 & 12.000 & 0.011 & 7 & 11.714 & 0.006 & 7 & 604920756091232256 \\
FS017-s96064 & 8:51:16.98 & 11:50:09.44 & 12.083 & 0.009 & 7 & 11.835 & 0.005 & 7 & 604920721731496064 \\
FS017-s02464 & 8:51:25.52 & 11:52:38.83 & 12.117 & 0.010 & 7 & 11.797 & 0.007 & 7 & 604921271487102464 \\
FS017-s82048 & 8:51:22.41 & 11:51:29.24 & 12.170 & 0.008 & 7 & 11.864 & 0.003 & 7 & 604921168408082048 \\
FS017-s33408 & 8:51:17.75 & 11:50:05.60 & 12.357 & 0.010 & 7 & 12.065 & 0.006 & 7 & 604920756091233408 \\
FS017 & 8:51:19.69 & 11:52:10.75 & 12.655 & 0.009 & 7 & 12.273 & 0.005 & 7 & 604921374566324992 \\
FS017-s82304 & 8:51:21.76 & 11:51:42.06 & 12.760 & 0.010 & 7 & 12.368 & 0.004 & 7 & 604921168408082304 \\
FS017-s49824 & 8:51:20.04 & 11:51:01.70 & 13.084 & 0.009 & 7 & 12.744 & 0.006 & 7 & 604921134048349824 \\
\enddata
\end{deluxetable*}

\newpage

\begin{figure*}[t]
\centering
\includegraphics[width=0.55\textwidth]{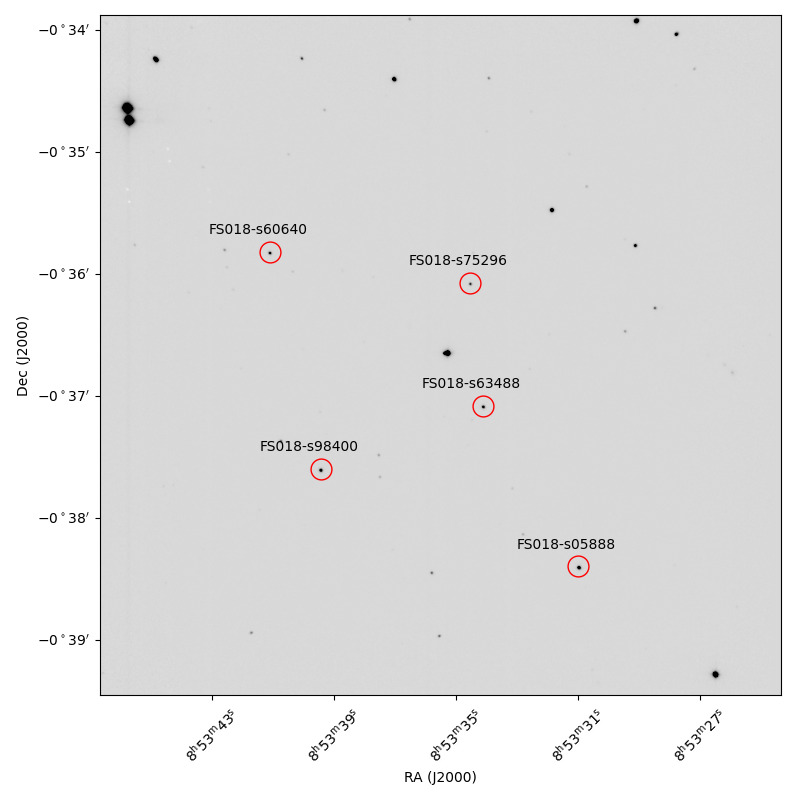} \\
\vspace{5pt}
\includegraphics[width=0.55\textwidth]{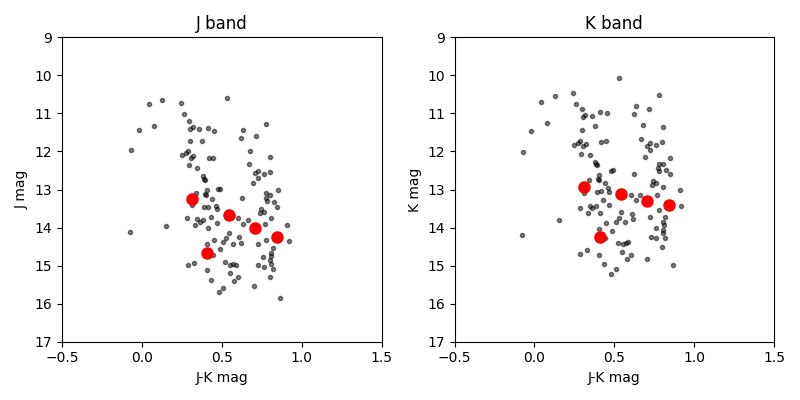}
\caption{FS018 field finding chart and color-magnitude diagrams}
\end{figure*}

\begin{deluxetable*}{ccc|ccc|ccc|c}[ht!]
\tabletypesize{\scriptsize}
\tablewidth{0pt} 
\tablecaption{FS018}
\tablehead{
\colhead{name} & \colhead{ra} & \colhead{dec} & 
\colhead{J} & \colhead{err J} & \colhead{epochs J} &
\colhead{K} & \colhead{err K} & \colhead{epochs K} &
\colhead{GAIA id}  
} 
\startdata 
FS018-s05888 & 8:53:31.19 & -0:38:26.70 & 13.247 & 0.007 & 8 & 12.934 & 0.004 & 8 & 3074350479674405888 \\
FS018-s98400 & 8:53:39.64 & -0:37:38.82 & 13.659 & 0.010 & 8 & 13.116 & 0.009 & 8 & 3074350926350998400 \\
FS018-s63488 & 8:53:34.29 & -0:37:07.72 & 14.000 & 0.005 & 8 & 13.292 & 0.006 & 8 & 3074350891991263488 \\
FS018-s60640 & 8:53:41.35 & -0:35:51.76 & 14.260 & 0.010 & 8 & 13.416 & 0.006 & 8 & 3074352506898960640 \\
FS018-s75296 & 8:53:34.75 & -0:36:07.33 & 14.669 & 0.010 & 8 & 14.260 & 0.008 & 8 & 3074353950007975296 \\
\enddata
\end{deluxetable*}

\newpage

\begin{figure*}[t]
\centering
\includegraphics[width=0.55\textwidth]{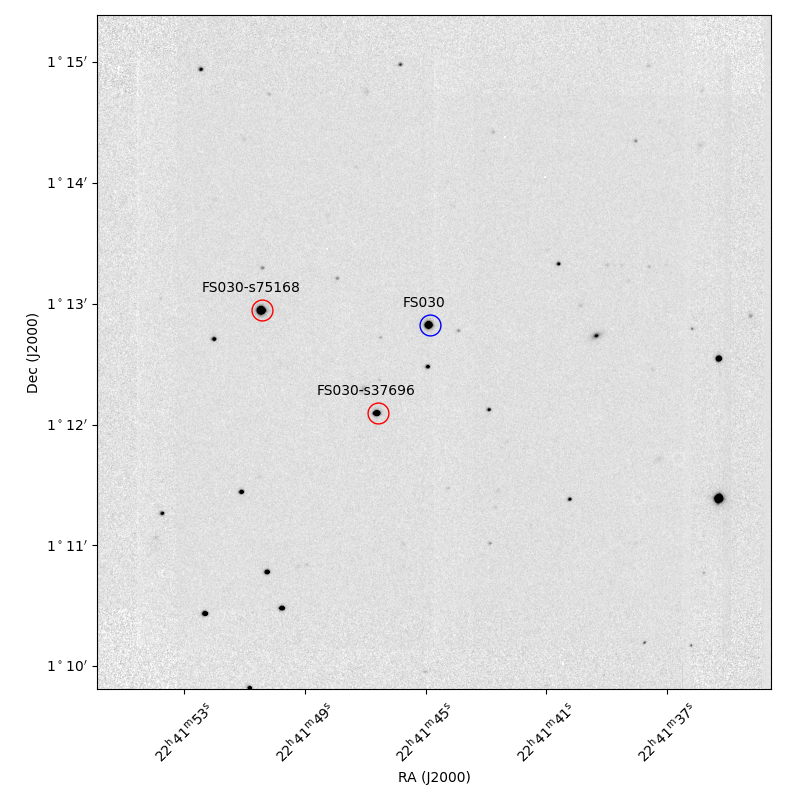} \\
\vspace{5pt}
\includegraphics[width=0.55\textwidth]{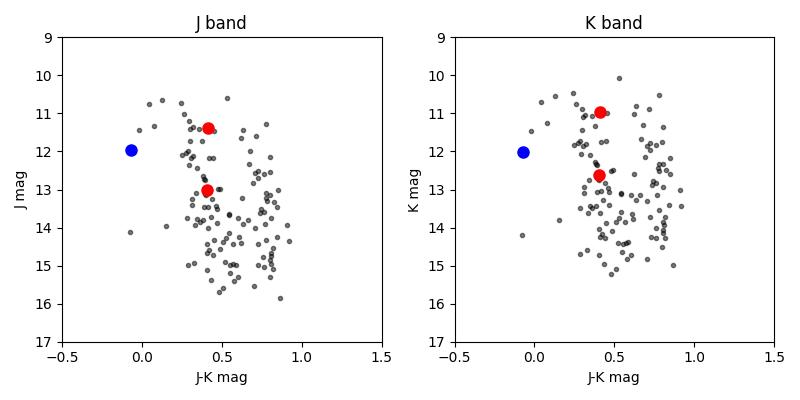}
\caption{FS030 field finding chart and color-magnitude diagrams}
\end{figure*}

\begin{deluxetable*}{ccc|ccc|ccc|c}[ht!]
\tabletypesize{\scriptsize}
\tablewidth{0pt} 
\tablecaption{FS030}
\tablehead{
\colhead{name} & \colhead{ra} & \colhead{dec} & 
\colhead{J} & \colhead{err J} & \colhead{epochs J} &
\colhead{K} & \colhead{err K} & \colhead{epochs K} &
\colhead{GAIA id}  
} 
\startdata 
FS030-s75168 & 22:41:50.24 & 1:12:43.25 & 11.383 & 0.011 & 6 & 10.972 & 0.007 & 6 & 2654543123279175168 \\
FS030 & 22:41:44.70 & 1:12:36.37 & 11.949 & 0.008 & 6 & 12.018 & 0.008 & 6 & 2654543161934285440 \\
FS030-s37696 & 22:41:46.40 & 1:11:52.20 & 13.018 & 0.007 & 6 & 12.613 & 0.004 & 6 & 2654543088919437696 \\
\enddata
\end{deluxetable*}

\newpage

\begin{figure*}[t]
\centering
\includegraphics[width=0.5\textwidth]{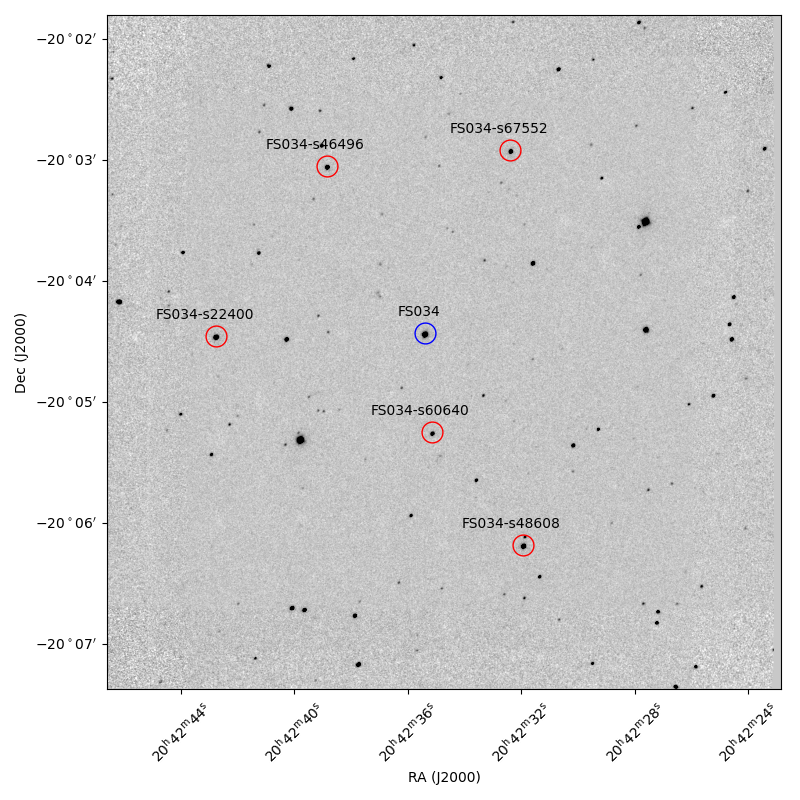} \\
\vspace{5pt}
\includegraphics[width=0.5\textwidth]{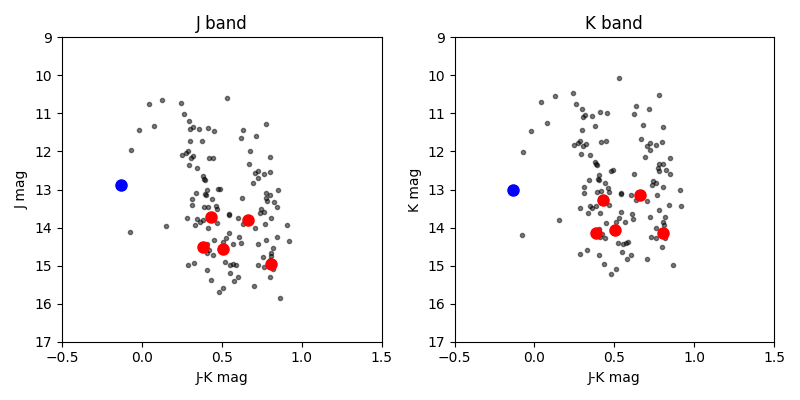}
\caption{FS034 field finding chart and color-magnitude diagrams}
\end{figure*}

\begin{deluxetable*}{ccc|ccc|ccc|c}[ht!]
\tabletypesize{\scriptsize}
\tablewidth{0pt} 
\tablecaption{FS034}
\tablehead{
\colhead{name} & \colhead{ra} & \colhead{dec} & 
\colhead{J} & \colhead{err J} & \colhead{epochs J} &
\colhead{K} & \colhead{err K} & \colhead{epochs K} &
\colhead{GAIA id}  
} 
\startdata 
FS034 & 20:42:34.75 & -20:04:35.93 & 12.872 & 0.009 & 11 & 13.000 & 0.009 & 11 & 6857939315643803776 \\ 
FS034-s22400 & 20:42:42.43 & -20:04:38.54 & 13.715 & 0.010 & 9 & 13.285 & 0.009 & 9 & 6857939624881622400 \\
FS034-s48608 & 20:42:31.61 & -20:06:22.27 & 13.812 & 0.009 & 11 & 13.149 & 0.008 & 11 & 6857939109486948608 \\
FS034-s46496 & 20:42:38.52 & -20:03:14.20 & 14.561 & 0.006 & 11 & 14.056 & 0.006 & 11 & 6857942682898346496 \\
FS034-s60640 & 20:42:34.81 & -20:05:26.37 & 14.948 & 0.009 & 10 & 14.140 & 0.008 & 10 & 6857939208267760640 \\
\enddata
\end{deluxetable*}

\newpage

\begin{figure*}[t]
\centering
\includegraphics[width=0.4\textwidth]{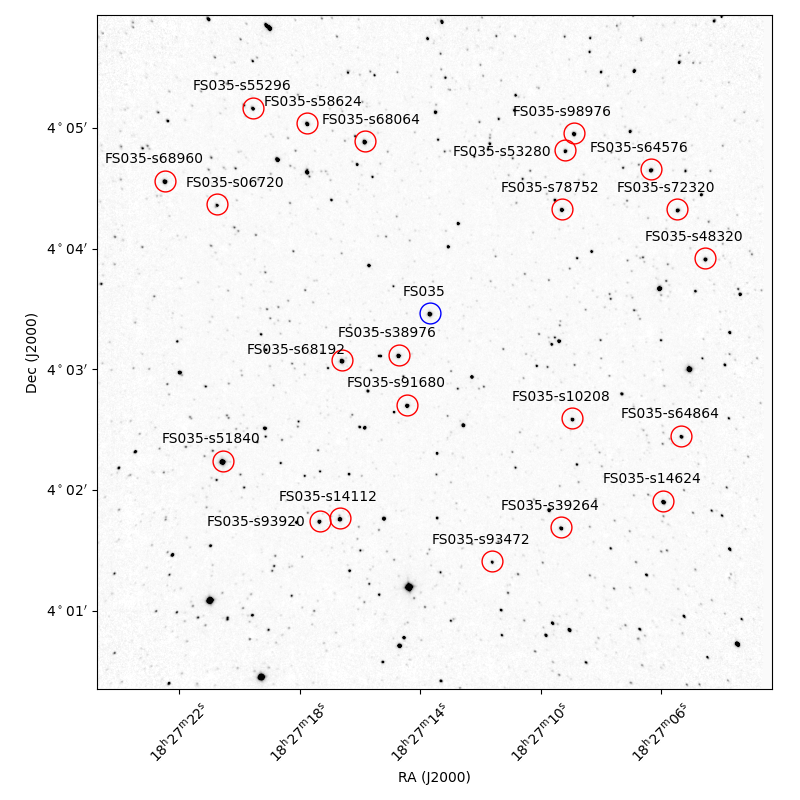} \\
\vspace{5pt}
\includegraphics[width=0.4\textwidth]{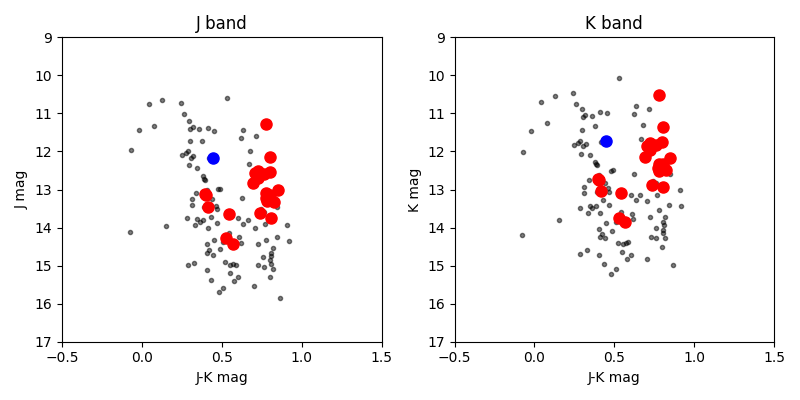}
\caption{FS035 field finding chart and color-magnitude diagrams}
\end{figure*}

\begin{deluxetable*}{ccc|ccc|ccc|c}[ht!]
\tabletypesize{\scriptsize}
\tablewidth{0pt} 
\tablecaption{FS035}
\tablehead{
\colhead{name} & \colhead{ra} & \colhead{dec} & 
\colhead{J} & \colhead{err J} & \colhead{epochs J} &
\colhead{K} & \colhead{err K} & \colhead{epochs K} &
\colhead{GAIA id}  
} 
\startdata 
FS035-s51840 & 18:27:20.35 & 4:01:56.29 & 11.289 & 0.007 & 6 & 10.511 & 0.008 & 6 & 4284122439177651840 \\
FS035-s68192 & 18:27:16.40 & 4:02:46.41 & 12.147 & 0.008 & 6 & 11.344 & 0.011 & 6 & 4284122709739768192 \\
FS035 & 18:27:13.50 & 4:03:09.80 & 12.182 & 0.007 & 6 & 11.734 & 0.009 & 6 & 4284122748415319936 \\
FS035-s14112 & 18:27:16.46 & 4:01:27.90 & 12.513 & 0.008 & 6 & 11.788 & 0.011 & 6 & 4284122370460714112 \\
FS035-s68960 & 18:27:22.27 & 4:04:15.46 & 12.550 & 0.008 & 6 & 11.749 & 0.011 & 6 & 4284128623930568960 \\
FS035-s68064 & 18:27:15.64 & 4:04:35.09 & 12.555 & 0.008 & 6 & 11.847 & 0.009 & 6 & 4284129414204568064 \\
FS035-s38976 & 18:27:14.54 & 4:02:49.07 & 12.604 & 0.008 & 6 & 11.841 & 0.010 & 6 & 4284122679695838976 \\
FS035-s14624 & 18:27:05.74 & 4:01:36.46 & 12.686 & 0.007 & 6 & 11.960 & 0.011 & 6 & 4284122610975914624 \\
FS035-s91680 & 18:27:14.23 & 4:02:24.39 & 12.830 & 0.008 & 6 & 12.135 & 0.010 & 6 & 4284122675379991680 \\
FS035-s58624 & 18:27:17.55 & 4:04:44.10 & 13.023 & 0.008 & 6 & 12.171 & 0.010 & 6 & 4284128692650058624 \\
FS035-s48320 & 18:27:04.34 & 4:03:36.90 & 13.097 & 0.006 & 6 & 12.317 & 0.010 & 6 & 4284123469969848320 \\
FS035-s78752 & 18:27:09.10 & 4:04:01.43 & 13.120 & 0.008 & 6 & 12.723 & 0.011 & 6 & 4284123504329578752 \\
FS035-s93920 & 18:27:17.15 & 4:01:26.72 & 13.143 & 0.008 & 6 & 12.340 & 0.010 & 6 & 4284122473537393920 \\
FS035-s64576 & 18:27:06.15 & 4:04:20.90 & 13.144 & 0.007 & 6 & 12.742 & 0.010 & 6 & 4284123573049064576 \\
FS035-s98976 & 18:27:08.70 & 4:04:39.06 & 13.222 & 0.008 & 6 & 12.446 & 0.011 & 6 & 4284123607408798976 \\
FS035-s39264 & 18:27:09.13 & 4:01:23.48 & 13.290 & 0.009 & 6 & 12.507 & 0.009 & 6 & 4284122537940939264 \\
FS035-s72320 & 18:27:05.27 & 4:04:01.24 & 13.322 & 0.007 & 6 & 12.498 & 0.009 & 6 & 4284123573049272320 \\
FS035-s10208 & 18:27:08.75 & 4:02:17.38 & 13.456 & 0.007 & 6 & 13.040 & 0.010 & 6 & 4284122645336110208 \\
FS035-s64864 & 18:27:05.15 & 4:02:08.85 & 13.610 & 0.009 & 6 & 12.872 & 0.011 & 6 & 4284123366890164864 \\
FS035-s53280 & 18:27:08.99 & 4:04:30.61 & 13.638 & 0.007 & 6 & 13.094 & 0.010 & 6 & 4284123603093153280 \\
FS035-s55296 & 18:27:19.35 & 4:04:51.71 & 13.749 & 0.008 & 6 & 12.941 & 0.009 & 6 & 4284128692650055296 \\
FS035-s06720 & 18:27:20.54 & 4:04:03.76 & 14.264 & 0.008 & 6 & 13.737 & 0.009 & 6 & 4284128623920206720 \\
FS035-s93472 & 18:27:11.41 & 4:01:06.61 & 14.424 & 0.007 & 6 & 13.856 & 0.007 & 6 & 4284122336097993472 \\
\enddata
\end{deluxetable*}

\newpage

\begin{figure*}[t]
\centering
\includegraphics[width=0.55\textwidth]{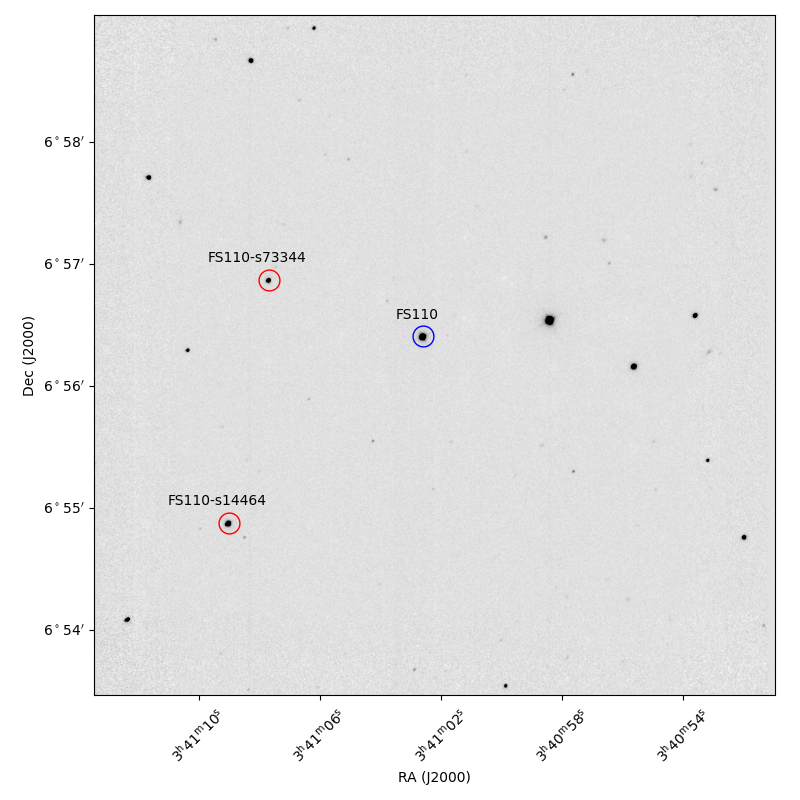} \\
\vspace{5pt}
\includegraphics[width=0.55\textwidth]{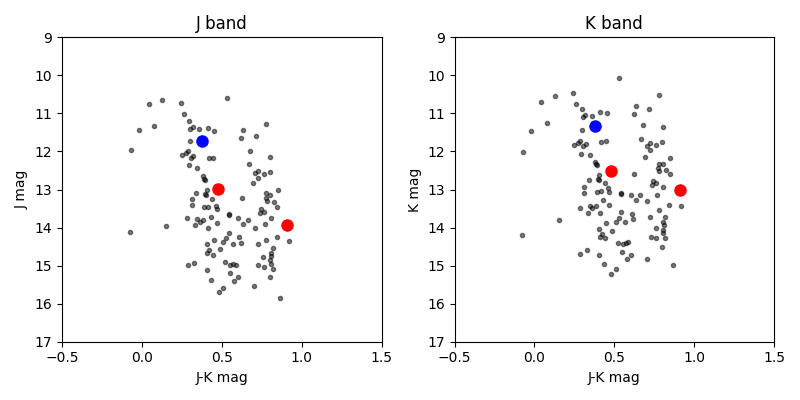}
\caption{FS110 field finding chart and color-magnitude diagrams}
\end{figure*}

\begin{deluxetable*}{ccc|ccc|ccc|c}[ht!]
\tabletypesize{\scriptsize}
\tablewidth{0pt} 
\tablecaption{FS110}
\tablehead{
\colhead{name} & \colhead{ra} & \colhead{dec} & 
\colhead{J} & \colhead{err J} & \colhead{epochs J} &
\colhead{K} & \colhead{err K} & \colhead{epochs K} &
\colhead{GAIA id}  
} 
\startdata 
FS110 & 3:41:02.22 & 6:56:16.43 & 11.715 & 0.007 & 11 & 11.336 & 0.004 & 11 & 3277706323464131968 \\
FS110-s14464 & 3:41:08.64 & 6:54:44.94 & 12.990 & 0.010 & 11 & 12.512 & 0.006 & 11 & 3277659113183614464 \\
FS110-s73344 & 3:41:07.29 & 6:56:44.49 & 13.928 & 0.006 & 11 & 13.018 & 0.003 & 11 & 3277659525500473344 \\
\enddata
\end{deluxetable*}

\newpage

\begin{figure*}[t]
\centering
\includegraphics[width=0.55\textwidth]{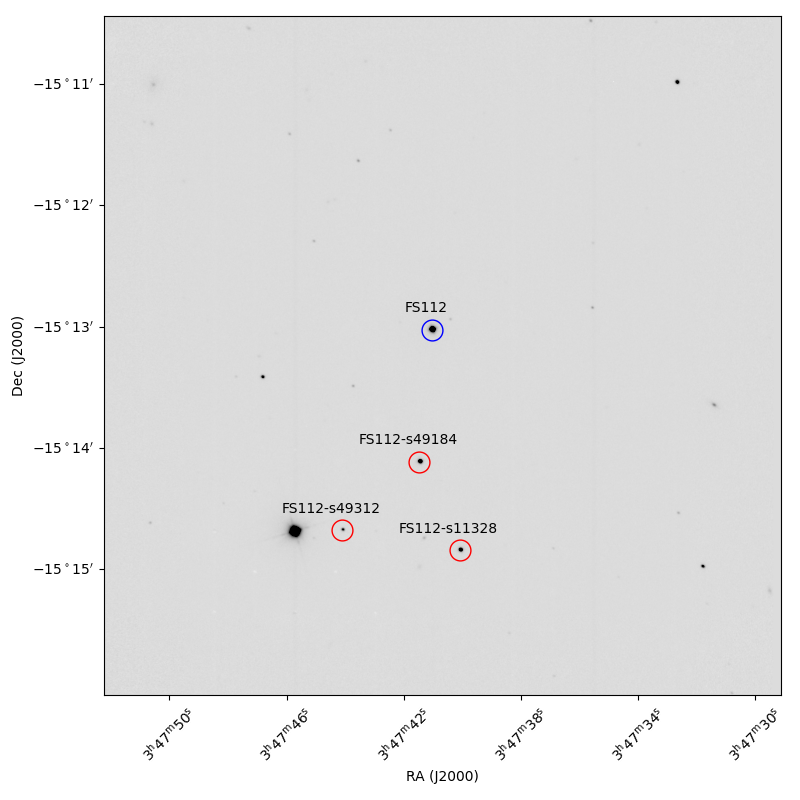} \\
\vspace{5pt}
\includegraphics[width=0.55\textwidth]{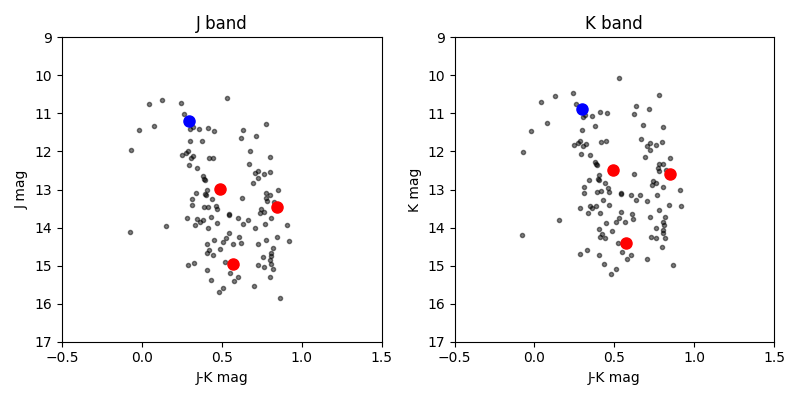}
\caption{FS112 field finding chart and color-magnitude diagrams}
\end{figure*}

\begin{deluxetable*}{ccc|ccc|ccc|c}[ht!]
\tabletypesize{\scriptsize}
\tablewidth{0pt} 
\tablecaption{FS112}
\tablehead{
\colhead{name} & \colhead{ra} & \colhead{dec} & 
\colhead{J} & \colhead{err J} & \colhead{epochs J} &
\colhead{K} & \colhead{err K} & \colhead{epochs K} &
\colhead{GAIA id}  
} 
\startdata 
FS112 & 3:47:40.72 & -15:13:14.59 & 11.190 & 0.007 & 10 & 10.893 & 0.005 & 10 & 5109048973678255488 \\
FS112-s49184 & 3:47:41.12 & -15:14:19.76 & 12.980 & 0.009 & 10 & 12.490 & 0.004 & 10 & 5109048698800349184 \\
FS112-s11328 & 3:47:39.72 & -15:15:03.20 & 13.447 & 0.008 & 10 & 12.600 & 0.004 & 10 & 5109048664440611328 \\
FS112-s49312 & 3:47:43.76 & -15:14:53.51 & 14.966 & 0.010 & 10 & 14.393 & 0.006 & 10 & 5109048698800349312 \\
\enddata
\end{deluxetable*}

\newpage

\begin{figure*}[t]
\centering
\includegraphics[width=0.55\textwidth]{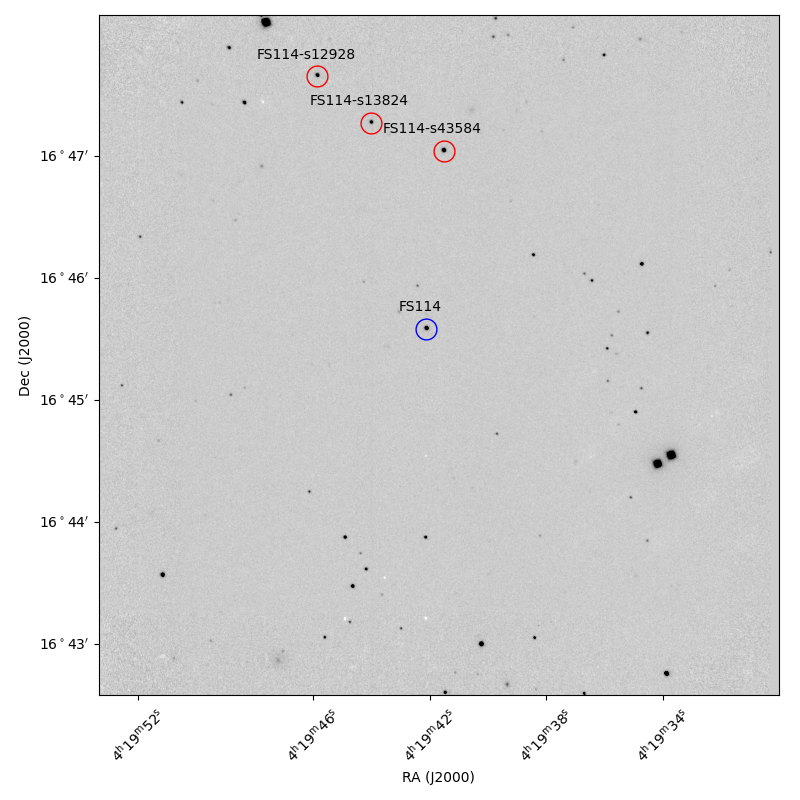} \\
\vspace{5pt}
\includegraphics[width=0.55\textwidth]{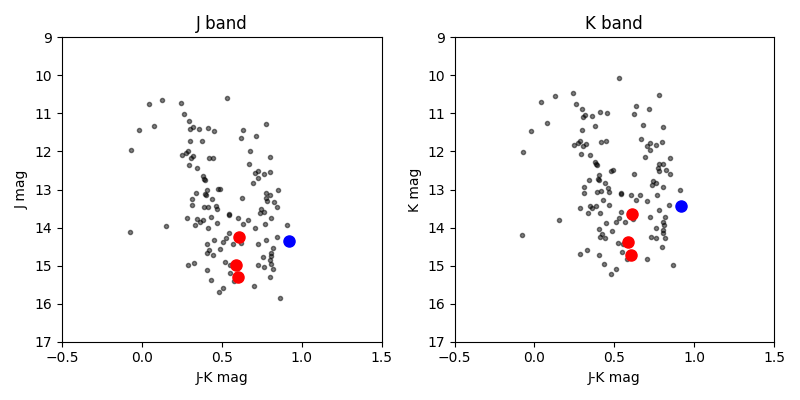}
\caption{FS114 field finding chart and color-magnitude diagrams}
\end{figure*}

\begin{deluxetable*}{ccc|ccc|ccc|c}[ht!]
\tabletypesize{\scriptsize}
\tablewidth{0pt} 
\tablecaption{FS114}
\tablehead{
\colhead{name} & \colhead{ra} & \colhead{dec} & 
\colhead{J} & \colhead{err J} & \colhead{epochs J} &
\colhead{K} & \colhead{err K} & \colhead{epochs K} &
\colhead{GAIA id}  
} 
\startdata 
FS114-s43584 & 4:19:41.20 & 16:46:48.45 & 14.246 & 0.006 & 23 & 13.635 & 0.004 & 23 & 3313880805773043584 \\
FS114 & 4:19:41.73 & 16:45:22.05 & 14.360 & 0.004 & 26 & 13.442 & 0.003 & 26 & 3313879946778443648 \\
FS114-s12928 & 4:19:45.53 & 16:47:25.32 & 14.981 & 0.008 & 17 & 14.392 & 0.010 & 17 & 3313880152938012928 \\
FS114-s13824 & 4:19:43.68 & 16:47:02.19 & 15.311 & 0.006 & 22 & 14.708 & 0.009 & 22 & 3313880152938013824 \\
\enddata
\end{deluxetable*}

\newpage

\begin{figure*}[t]
\centering
\includegraphics[width=0.45\textwidth]{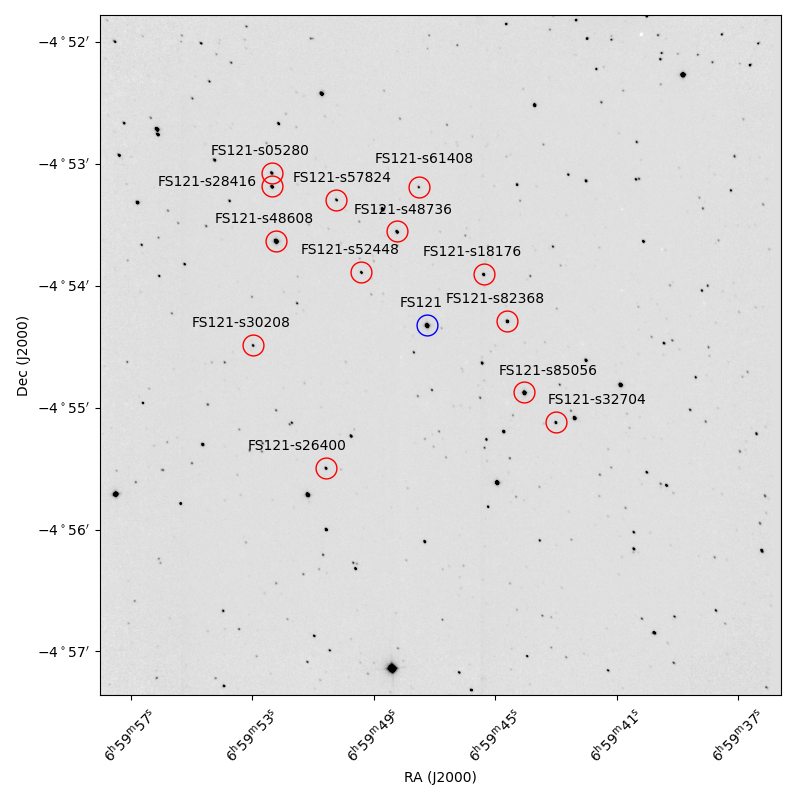} \\
\vspace{5pt}
\includegraphics[width=0.45\textwidth]{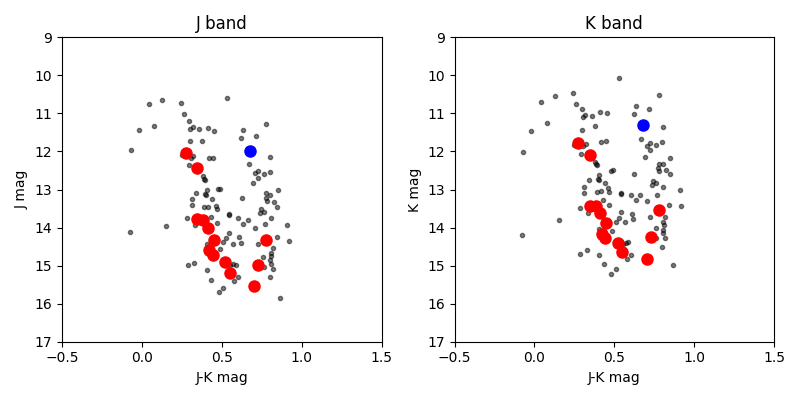}
\caption{FS121 field finding chart and color-magnitude diagrams}
\end{figure*}

\begin{deluxetable*}{ccc|ccc|ccc|c}[ht!]
\tabletypesize{\scriptsize}
\tablewidth{0pt} 
\tablecaption{FS121}
\tablehead{
\colhead{name} & \colhead{ra} & \colhead{dec} & 
\colhead{J} & \colhead{err J} & \colhead{epochs J} &
\colhead{K} & \colhead{err K} & \colhead{epochs K} &
\colhead{GAIA id}  
} 
\startdata 
FS121 & 6:59:46.77 & -4:54:33.67 & 11.977 & 0.003 & 16 & 11.300 & 0.003 & 16 & 3101625583593341568 \\
FS121-s48608 & 6:59:51.74 & -4:53:52.27 & 12.051 & 0.006 & 16 & 11.775 & 0.006 & 16 & 3101625686672548608 \\
FS121-s85056 & 6:59:43.57 & -4:55:06.81 & 12.434 & 0.006 & 16 & 12.087 & 0.005 & 16 & 3101625617953085056 \\
FS121-s28416 & 6:59:51.88 & -4:53:25.54 & 13.777 & 0.007 & 16 & 13.429 & 0.009 & 16 & 3101625716732828416 \\
FS121-s82368 & 6:59:44.13 & -4:54:31.69 & 13.811 & 0.004 & 16 & 13.426 & 0.005 & 16 & 3101625617953082368 \\
FS121-s48736 & 6:59:47.77 & -4:53:47.64 & 14.015 & 0.005 & 16 & 13.604 & 0.006 & 16 & 3101625751092548736 \\
FS121-s18176 & 6:59:44.92 & -4:54:08.63 & 14.323 & 0.005 & 16 & 13.543 & 0.006 & 16 & 3101625652312818176 \\
FS121-s05280 & 6:59:51.89 & -4:53:18.68 & 14.328 & 0.009 & 16 & 13.877 & 0.009 & 16 & 3101625785452305280 \\
FS121-s32704 & 6:59:42.54 & -4:55:21.41 & 14.581 & 0.006 & 16 & 14.159 & 0.006 & 16 & 3101602145952232704 \\
FS121-s26400 & 6:59:50.11 & -4:55:43.84 & 14.721 & 0.008 & 16 & 14.277 & 0.005 & 16 & 3101613793903526400 \\
FS121-s52448 & 6:59:48.94 & -4:54:07.64 & 14.917 & 0.004 & 16 & 14.396 & 0.009 & 16 & 3101625686672552448 \\
FS121-s57824 & 6:59:49.76 & -4:53:32.07 & 14.986 & 0.007 & 16 & 14.259 & 0.009 & 16 & 3101625751092557824 \\
FS121-s30208 & 6:59:52.50 & -4:54:43.50 & 15.184 & 0.005 & 16 & 14.633 & 0.009 & 16 & 3101613935642030208 \\
FS121-s61408 & 6:59:47.05 & -4:53:25.76 & 15.538 & 0.007 & 16 & 14.835 & 0.011 & 16 & 3101625751092561408 \\
\enddata
\end{deluxetable*}

\newpage

\begin{figure*}[t]
\centering
\includegraphics[width=0.55\textwidth]{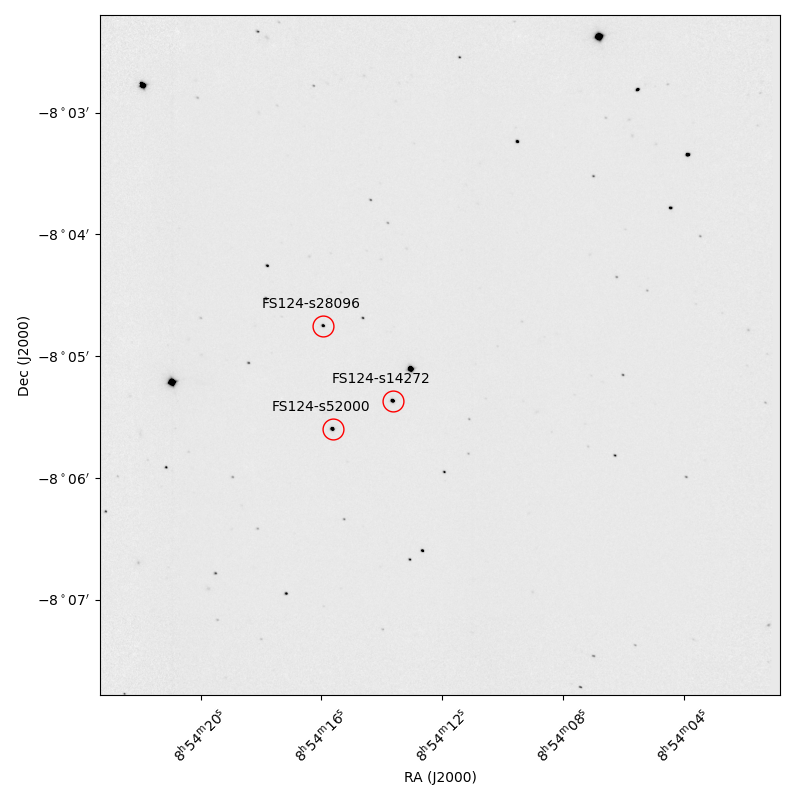} \\
\vspace{5pt}
\includegraphics[width=0.55\textwidth]{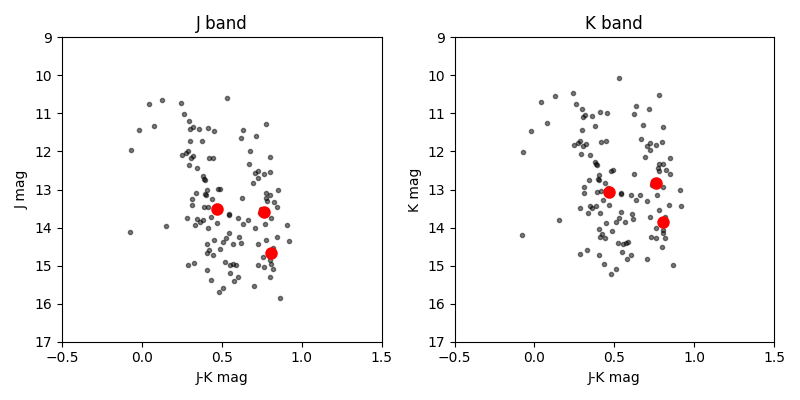}
\caption{FS124 field finding chart and color-magnitude diagrams}
\end{figure*}

\begin{deluxetable*}{ccc|ccc|ccc|c}[ht!]
\tabletypesize{\scriptsize}
\tablewidth{0pt} 
\tablecaption{FS124}
\tablehead{
\colhead{name} & \colhead{ra} & \colhead{dec} & 
\colhead{J} & \colhead{err J} & \colhead{epochs J} &
\colhead{K} & \colhead{err K} & \colhead{epochs K} &
\colhead{GAIA id}  
} 
\startdata 
FS124-s14272 & 8:54:13.83 & -8:05:27.00 & 13.524 & 0.007 & 14 & 13.056 & 0.005 & 14 & 5756746672027014272 \\
FS124-s52000 & 8:54:15.83 & -8:05:41.04 & 13.584 & 0.008 & 14 & 12.820 & 0.005 & 14 & 5756746706386752000 \\
FS124-s28096 & 8:54:16.12 & -8:04:49.83 & 14.666 & 0.005 & 14 & 13.860 & 0.004 & 14 & 5756746775106228096 \\
\enddata
\end{deluxetable*}

\newpage

\begin{figure*}[t]
\centering
\includegraphics[width=0.55\textwidth]{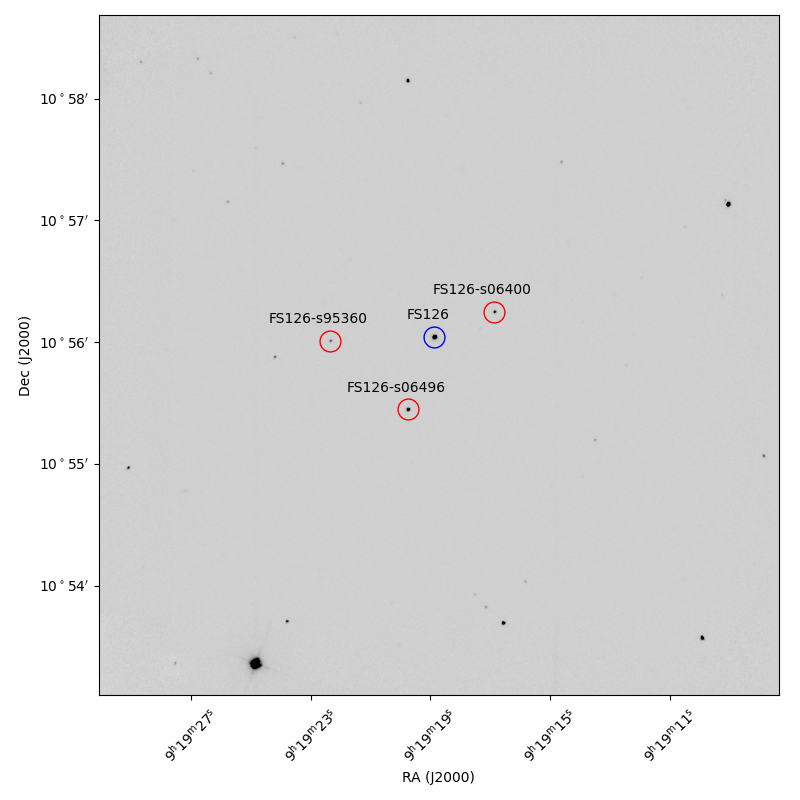} \\
\vspace{5pt}
\includegraphics[width=0.55\textwidth]{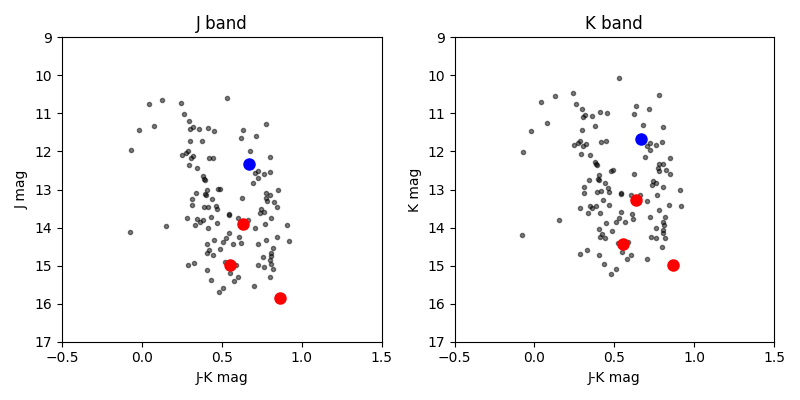}
\caption{FS126 field finding chart and color-magnitude diagrams}
\end{figure*}

\begin{deluxetable*}{ccc|ccc|ccc|c}[ht!]
\tabletypesize{\scriptsize}
\tablewidth{0pt} 
\tablecaption{FS126}
\tablehead{
\colhead{name} & \colhead{ra} & \colhead{dec} & 
\colhead{J} & \colhead{err J} & \colhead{epochs J} &
\colhead{K} & \colhead{err K} & \colhead{epochs K} &
\colhead{GAIA id}  
} 
\startdata 
FS126 & 9:19:18.75 & 10:55:51.10 & 12.330 & 0.009 & 9 & 11.662 & 0.006 & 9 & 592615193750958464 \\
FS126-s06496 & 9:19:19.62 & 10:55:15.38 & 13.901 & 0.004 & 9 & 13.268 & 0.005 & 9 & 592615086376506496 \\
FS126-s06400 & 9:19:16.73 & 10:56:03.41 & 14.993 & 0.011 & 9 & 14.441 & 0.009 & 9 & 592615258175206400 \\
FS126-s95360 & 9:19:22.20 & 10:55:49.09 & 15.854 & 0.010 & 9 & 14.988 & 0.010 & 9 & 592615193751295360 \\
\enddata
\end{deluxetable*}

\newpage

\begin{figure*}[t]
\centering
\includegraphics[width=0.5\textwidth]{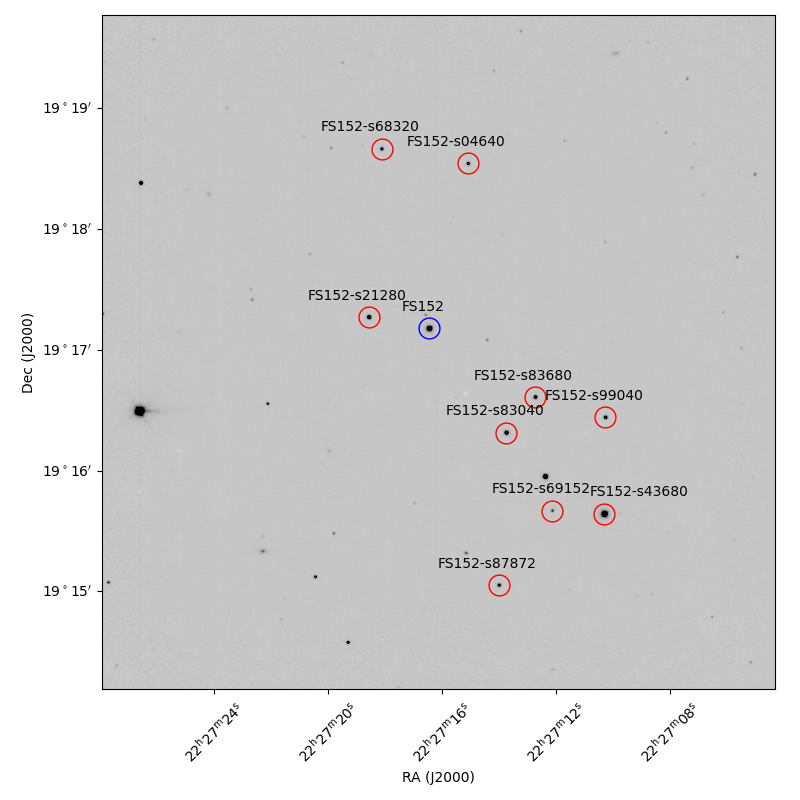} \\
\vspace{5pt}
\includegraphics[width=0.5\textwidth]{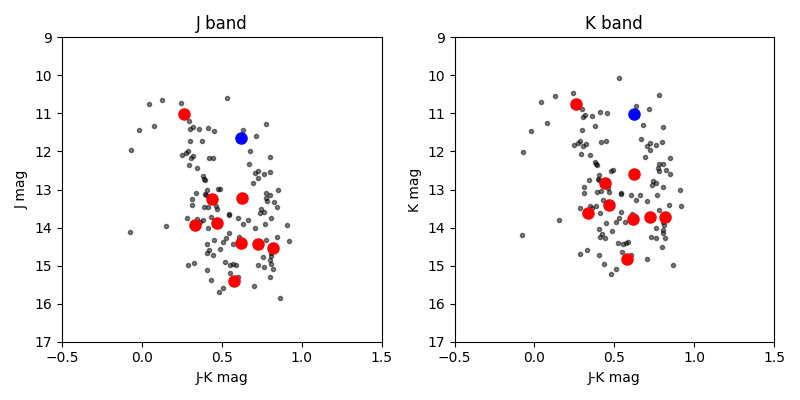}
\caption{FS152 field finding chart and color-magnitude diagrams}
\end{figure*}

\begin{deluxetable*}{ccc|ccc|ccc|c}[ht!]
\tabletypesize{\scriptsize}
\tablewidth{0pt} 
\tablecaption{FS152}
\tablehead{
\colhead{name} & \colhead{ra} & \colhead{dec} & 
\colhead{J} & \colhead{err J} & \colhead{epochs J} &
\colhead{K} & \colhead{err K} & \colhead{epochs K} &
\colhead{GAIA id}  
} 
\startdata 
FS152-s43680 & 22:27:10.01 & 19:15:23.11 & 11.017 & 0.006 & 11 & 10.756 & 0.006 & 11 & 1777401846705943680 \\
FS152 & 22:27:16.14 & 19:16:55.41 & 11.639 & 0.004 & 11 & 11.017 & 0.007 & 11 & 1777402018504634496 \\
FS152-s21280 & 22:27:18.26 & 19:17:00.86 & 13.218 & 0.005 & 11 & 12.592 & 0.008 & 11 & 1777402022799921280 \\
FS152-s83040 & 22:27:13.43 & 19:16:03.40 & 13.262 & 0.005 & 11 & 12.820 & 0.008 & 11 & 1777401988440183040 \\
FS152-s83680 & 22:27:12.41 & 19:16:21.06 & 13.872 & 0.004 & 11 & 13.404 & 0.005 & 11 & 1777401988440183680 \\
FS152-s99040 & 22:27:09.95 & 19:16:11.06 & 13.945 & 0.008 & 11 & 13.610 & 0.007 & 11 & 1777402091519399040 \\
FS152-s87872 & 22:27:13.69 & 19:14:47.74 & 14.395 & 0.010 & 11 & 13.776 & 0.011 & 11 & 1777401816641487872 \\
FS152-s68320 & 22:27:17.80 & 19:18:24.51 & 14.442 & 0.005 & 11 & 13.718 & 0.007 & 11 & 1777403702132168320 \\
FS152-s04640 & 22:27:14.77 & 19:18:17.29 & 14.531 & 0.006 & 11 & 13.713 & 0.006 & 11 & 1777403706427104640 \\
FS152-s69152 & 22:27:11.83 & 19:15:24.69 & 15.392 & 0.009 & 11 & 14.814 & 0.009 & 11 & 1777401881065969152 \\
\enddata
\end{deluxetable*}

\newpage

\begin{figure*}[t]
\centering
\includegraphics[width=0.55\textwidth]{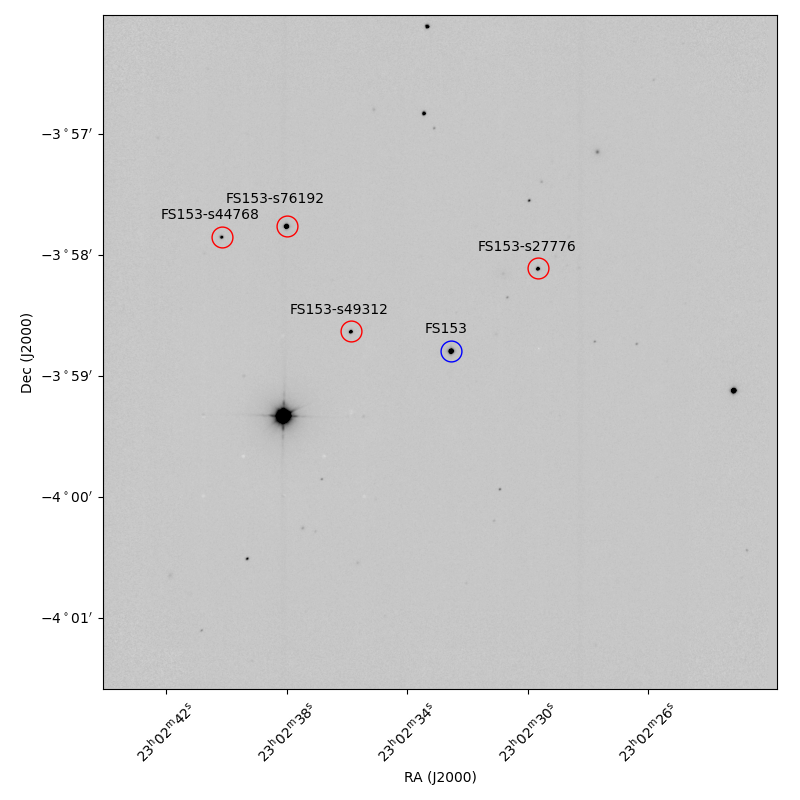} \\
\vspace{5pt}
\includegraphics[width=0.55\textwidth]{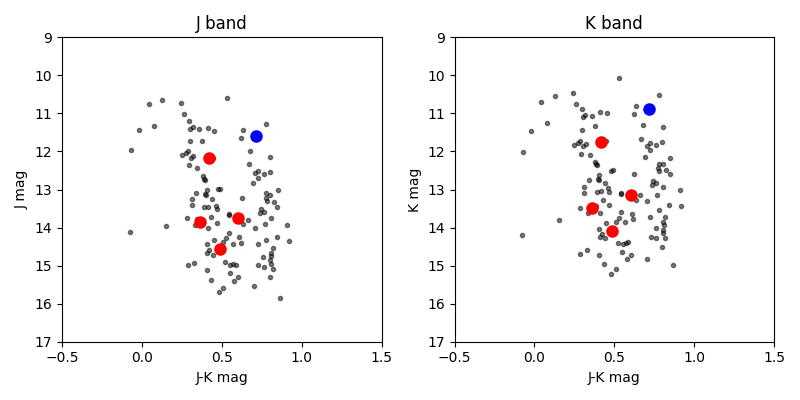}
\caption{FS153 field finding chart and color-magnitude diagrams}
\end{figure*}

\begin{deluxetable*}{ccc|ccc|ccc|c}[ht!]
\tabletypesize{\scriptsize}
\tablewidth{0pt} 
\tablecaption{FS153}
\tablehead{
\colhead{name} & \colhead{ra} & \colhead{dec} & 
\colhead{J} & \colhead{err J} & \colhead{epochs J} &
\colhead{K} & \colhead{err K} & \colhead{epochs K} &
\colhead{GAIA id}  
} 
\startdata 
FS153 & 23:02:32.08 & -3:58:53.03 & 11.590 & 0.005 & 13 & 10.874 & 0.006 & 13 & 2636398540016611200 \\
FS153-s76192 & 23:02:37.53 & -3:57:51.46 & 12.180 & 0.006 & 13 & 11.763 & 0.006 & 13 & 2636398677455576192 \\
FS153-s49312 & 23:02:35.40 & -3:58:43.55 & 13.740 & 0.005 & 13 & 13.137 & 0.006 & 13 & 2636398574376349312 \\
FS153-s27776 & 23:02:29.20 & -3:58:12.29 & 13.841 & 0.008 & 13 & 13.480 & 0.008 & 13 & 2636398604440827776 \\
FS153-s44768 & 23:02:39.68 & -3:57:56.67 & 14.566 & 0.006 & 13 & 14.079 & 0.009 & 13 & 2636395752582544768 \\
\enddata
\end{deluxetable*}

\newpage

\begin{figure*}[t]
\centering
\includegraphics[width=0.55\textwidth]{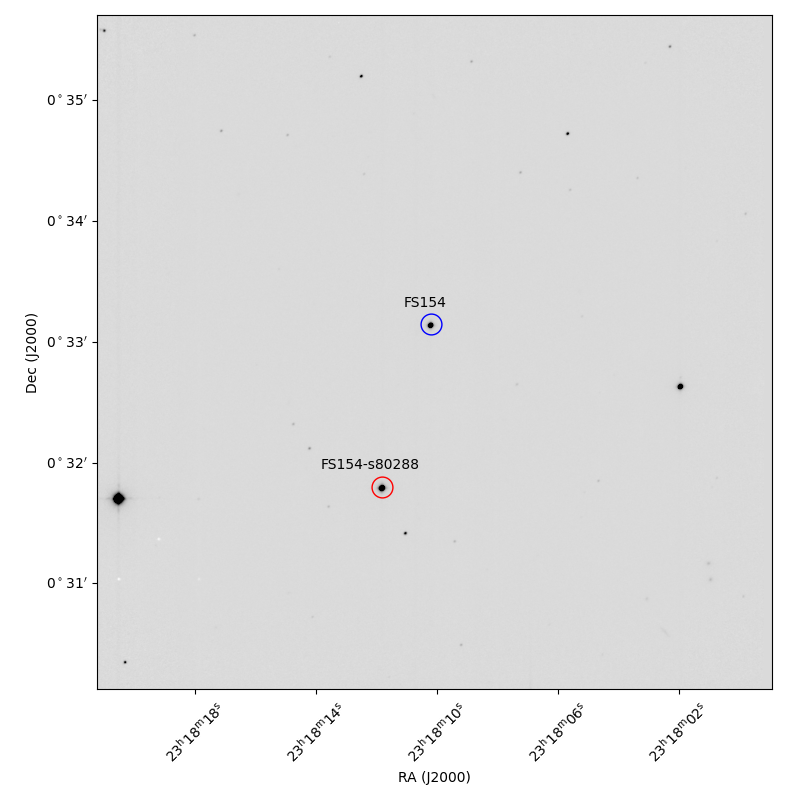} \\
\vspace{5pt}
\includegraphics[width=0.55\textwidth]{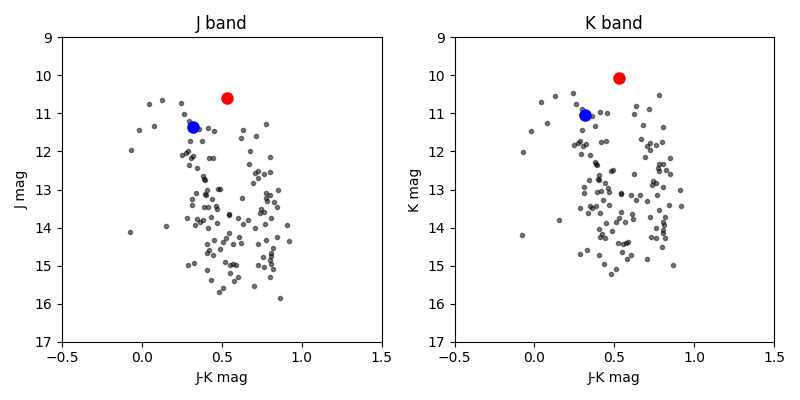}
\caption{FS154 field finding chart and color-magnitude diagrams}
\end{figure*}

\begin{deluxetable*}{ccc|ccc|ccc|c}[ht!]
\tabletypesize{\scriptsize}
\tablewidth{0pt} 
\tablecaption{FS154}
\tablehead{
\colhead{name} & \colhead{ra} & \colhead{dec} & 
\colhead{J} & \colhead{err J} & \colhead{epochs J} &
\colhead{K} & \colhead{err K} & \colhead{epochs K} &
\colhead{GAIA id}  
} 
\startdata 
FS154-s80288 & 23:18:11.63 & 0:31:35.57 & 10.597 & 0.006 & 9 & 10.065 & 0.010 & 9 & 2645250501973180288 \\
FS154 & 23:18:10.02 & 0:32:56.09 & 11.356 & 0.003 & 9 & 11.038 & 0.004 & 9 & 2645253559989894912 \\
\enddata
\end{deluxetable*}




\end{document}